\documentclass[11pt]{spieman}  

\usepackage{amsmath,amsfonts,amssymb}
\usepackage{graphicx}
\usepackage{setspace}
\usepackage{lineno}   
\usepackage{booktabs}
\usepackage{float}

\graphicspath{{tripetal_paper_figs/}}

\hypersetup{colorlinks=true, linkcolor=blue, citecolor=blue, urlcolor=blue}

\newcommand{\degsq}{\mathrm{deg}^2}

\title{Optimizing Field Geometry for Modular Micropore-Optics Soft X-ray Time-Domain Surveys}

\author[a,*]{Nicholas E. White}
\author[b,*]{Massimiliano Galeazzi}
\author[b]{Russell Roberts}
\author[c]{Kip D. Kuntz}
\affil[a]{The George Washington University, Washington, DC, USA}
\affil[b]{Department of Physics, University of Miami, 1320 Campo Sano Dr., Coral Gables, FL 33146, USA}
\affil[c]{Department of Physics and Astronomy, Johns Hopkins University, 3701 San Martin Drive, Baltimore, MD 21218, USA}

\begin{document}
\maketitle

\begin{abstract}
Wide-field micropore (lobster eye) optics enable focusing soft X-ray instruments with fields of view of hundreds of square degrees, but square-format modules require careful geometric arrangement to support both wide-area surveys and sensitive pointed observations. We examine this design problem using the X-ray TRansient Array / Micropore Optics X-ray Imager trade study, developed for a NASA Small Explorer mission concept, as a worked example. Single-module performance scaling is combined with geometric survey simulations to compare modular field layouts, overlap regions, and repeated survey tilings. For realistic micropore-optic parameters, increasing module size primarily expands the field of view once the boresight-centered effective area saturates, making modular architecture and field geometry central design variables. A three-module tri-petal configuration, formed by differentially rolling and offsetting three 10° × 10° square fields, produces a compound field of approximately 240 deg² with one-module discovery regions, two-module bridge regions, and a compact three-module core. When repeated in a dense hexagonal lattice, the footprint closes gaps, builds a more uniform exposure floor, and provides enhanced-depth survey regions. This geometry offers an efficient compromise among survey coverage, pointed sensitivity, system complexity, and resilience, and can be scaled to larger persistent soft X-ray monitoring architectures.
\end{abstract}


{\noindent \footnotesize\textbf{*}Co-corresponding authors: Nicholas E. White, newhite@gwu.edu;
Massimiliano Galeazzi, galeazzi@miami.edu}

\begin{spacing}{1}    

\section{Introduction}

The time-domain X-ray sky is rich in transient and highly variable phenomena spanning a wide range of spatial, temporal, and energetic scales, including gamma-ray bursts, tidal disruption events, supernova shock breakouts, magnetar flares, X-ray counterparts of gravitational wave sources, and accretion-powered outbursts from compact binaries
\cite{Willingale2017GRBFast,Gezari2021TDE,Waxman2017ShockBreakout,Kaspi2017Magnetars,Abbott2017GW170817GRB,Troja2017GW170817Xray,Remillard2006XRB}.
Many of these events are unpredictable and short-lived, with emission that can peak in the soft X-ray band (0.2-5.0 keV), making their discovery and characterization critically dependent on instruments with both a large instantaneous field of view and sufficient sensitivity \cite{Yuan2022}.

Swift provides the appropriate reference point for a modern high-energy transient mission. Its observing strategy is fundamentally reactive: an event trigger is followed by rapid slews of narrow-field focusing telescopes for X-ray and UV/optical afterglow observations \cite{Gehrels2004Swift,Burrows2005XRT}. The resulting soft X-ray sky coverage is extremely valuable and has generated deep serendipitous XRT source catalogues
\cite{Evans2014SwiftXRT1SXPS}, but it is an opportunistic archive rather than a designed survey. The accumulated exposure pattern is set by the history of GRBs, ToOs, orbital visibility, and scheduling constraints, so Swift does not tile a prescribed sky region on a controlled lattice or maintain a planned soft X-ray depth and revisit pattern. Further, Swift does not use its soft X-ray telescope to discover transients; it relies on the BAT and on other telescopes to discover transients.

The observing problem addressed here is different. Current time-domain and multimessenger astronomy requires both discovery of soft X-ray transients and rapid follow-up of events found at other wavelengths or messengers. The relevant localization regions span a very large range: arcsecond-scale positions can be followed efficiently with conventional narrow-field Wolter-I telescopes, whereas gravitational-wave, neutrino, poorly localized gamma-ray, and fast-transient alerts can cover hundreds to thousands of square degrees. The latter case requires not only a large instantaneous field of view, but also a survey geometry that can mosaic large probability regions quickly while producing a controlled and relatively uniform exposure map.

Einstein Probe has demonstrated the power of mission-scale wide-field focusing soft X-ray monitoring \cite{Yuan2022}. The trade studied here asks how that strategy can be extended to fainter transients while retaining large sky coverage and useful pointed-follow-up capability. 
The key distinction from both Swift and EP is therefore not simply field of view or total sky area accumulated over time, but control of the observing geometry; the ability to survey large areas to detect transients within the soft X-ray window, the ability to mosaic over large error boxes to find transients detected by other telescopes, as well as enhanced-depth follow-up with the {\it same} focusing soft X-ray instrument.

Micropore optics (MPOs), often called lobster-eye optics because of their analogy to crustacean eyes, provide a practical way to combine focusing soft X-ray sensitivity with very large instantaneous sky coverage. By concentrating source photons into an arcminute-scale imaging response while covering large solid angles, lobster-eye telescopes increase grasp and improve signal-to-noise relative to non-focusing monitors, while retaining positions useful for counterpart association and follow-up.

The lobster-eye X-ray concept was introduced by Angel \cite{Angel1979}, with related wide-field focusing concepts and micropore implementations developed in subsequent theoretical and experimental work
\cite{Schmidt1975,Fraser1992,Kaaret1992,Hudec2018,Brunton97,Priedhorsky1996ExtraordinaryASM,Fraser02}. MPOs are composed of many square micropores, closely packed together in a lattice structure, with the walls of the micropores normal to the surface of a sphere. X-ray focusing is achieved by internal grazing incidence reflection off orthogonal walls within the micropore, resulting in a focal surface at half the radius of curvature of the sphere. For an ideal optic, the intrinsic angular resolution is limited only by spherical aberration and geometric projection of the micropore ($\sim$ 30 arcseconds \cite{Willingale16}).

MPOs have been successfully qualified and flown for the first time onboard two sounding rockets (36.283-DXL/STORM \& 36.305-DXL/CuPID \cite{Collier15, Thomas13, Galeazzi14}), and have been space qualified and launched for the MIXS instruments on BepiColombo \cite{Huovelin20}, the CuPID cubesat \cite{Walsh21}, and the Lunar Environment Heliospheric X-ray Imager (LEXI \cite{Walsh24}). The Lobster Eye Imager for Astronomy (LEIA) provided the first in-orbit demonstration of wide-field focusing soft X-ray imaging with lobster-eye MPOs \cite{Zhang2022LEIA,Ling2023LEIA}.  Einstein Probe (EP) provides the current mission-scale reference point for modular lobster-eye soft X-ray monitoring.  The EP Wide-field X-ray Telescope (WXT) demonstrates how an array of MPO modules can deliver very large (3850 square degree) instantaneous sky coverage for high-cadence (every three orbits) soft X-ray transient discovery, while the EP Follow-up X-ray Telescope (EP/FXT) provides deeper pointed follow-up after a trigger \cite{Yuan2022}. However, such a mission requires two instruments for a single energy band.  Can a compact set of square-format MPO modules be arranged to provide {\it both} a wide field survey like the EP WXT and enhanced-depth targeted observations, like the EP/FXT, in order to allow a second instrument to cover a different energy band?

The X-ray TRansient Array (XTRA) \cite{White2026XTRA_SPIE} Micropore Optics X-ray Imager (MOXI) trade study addresses this question for a Small Explorer-scale time-domain mission, assumed to be in an L2 orbit. MPO modules offer a larger flexibility in mission design compared to traditional Wolter type I X-ray telescopes. Shock Interaction and Breakout EXplorer (SIBEX) concept considered the potential advantages and flexibility of a modular design \cite{Roming21} while the Lobster-eye X-ray Telescope (LXT) is implementing a compact design to be launched on a sounding rocket next year \cite{Galeazzi26}. 
The particular tri-petal configuration studied here occupies the intermediate regime between fully co-pointed modules, which maximize depth over a small field, and fully separated modules, which maximize geometric area but provide no enhanced-depth region.

This paper follows the design logic captured in the XTRA/MOXI survey-configuration trade study.  The relevant requirements are to exceed roughly \(200\ \degsq\) for GRB and multimessenger follow-up, maintain a sensitive and uniform large-area survey, preserve a compact overlap region for pointed observations, and minimize the number of telescope modules. 

We first review the single-module MPO performance trades that set the design space, including effective area, focal length, detector format, angular resolution, vignetting, and the saturation of boresight-centered effective area with increasing module size.  We then discuss the system-level trade between monolithic and modular architectures. In that context, XTRA/MOXI is used as the worked example: three square-format modules are differentially rolled and offset to form a tri-petal footprint with broad single-module petals, two-module bridges, and a compact three-module core for targeted pointed observations.  We show how this \(\sim241\ \degsq\) compound field interlocks on a \(10^\circ\times8.6^\circ\) hexagonal-knit survey lattice, compare it with co-pointed, separated, no-rotation, and Cartesian alternatives.  Finally, we consider how the same tri-petal principle could be scaled from an Explorer-class survey instrument to persistent, one-third-sky soft X-ray monitor arrays.

\section{Single Module Performance}

The performance of a single micropore optics (MPO) module is determined by a combination of geometric and materials parameters that define its effective area, Field of View (FoV), point spread function (PSF) and effective energy range. The quantities of primary relevance for instrument design are: (1) the effective area at the boresight (the center of the FoV), (2) vignetting or the effective area away from the boresight, (3) the total FoV, and (4) the accessible energy band. These quantities jointly determine the grasp of the instrument and its sensitivity to transient and diffuse emission.
For MPO modules, several physical parameters can be adjusted to optimize performance, including the focal length, the pore-wall coating, the optic size, and the pore aspect ratio (defined as the ratio of pore length to pore size). In practice, the available parameter space is constrained by manufacturing considerations and existing hardware. In this work, we adopt commercially available square MPO facets (e.g., Photonis \cite{Benzaid2026} or equivalent) with dimensions of 4 cm × 4 cm, a pore aspect ratio of 50, and iridium coating. Under these assumptions, the optimization reduces primarily to selecting the focal length and the number of MPO facets that must be mosaicked to form a module. It should be noted that the focal plane detector should have a linear dimension that is roughly half of that of the optic in order to fully cover the region without strong vignetting.

We do not consider designs in which the pore aspect ratio \(L/D\) varies across
the optic.  Such designs can be useful for increasing the boresight effective
area of instruments optimized for pointed imaging, because outer facets can be
made to contribute more efficiently to a source near the field center.
They are less appropriate for the present survey-geometry problem, where the
goal is a broad detector field with relatively uniform response rather than
maximum response at a single direction.

\begin{figure*}[h]
\centering
\includegraphics[width=0.35\textwidth]{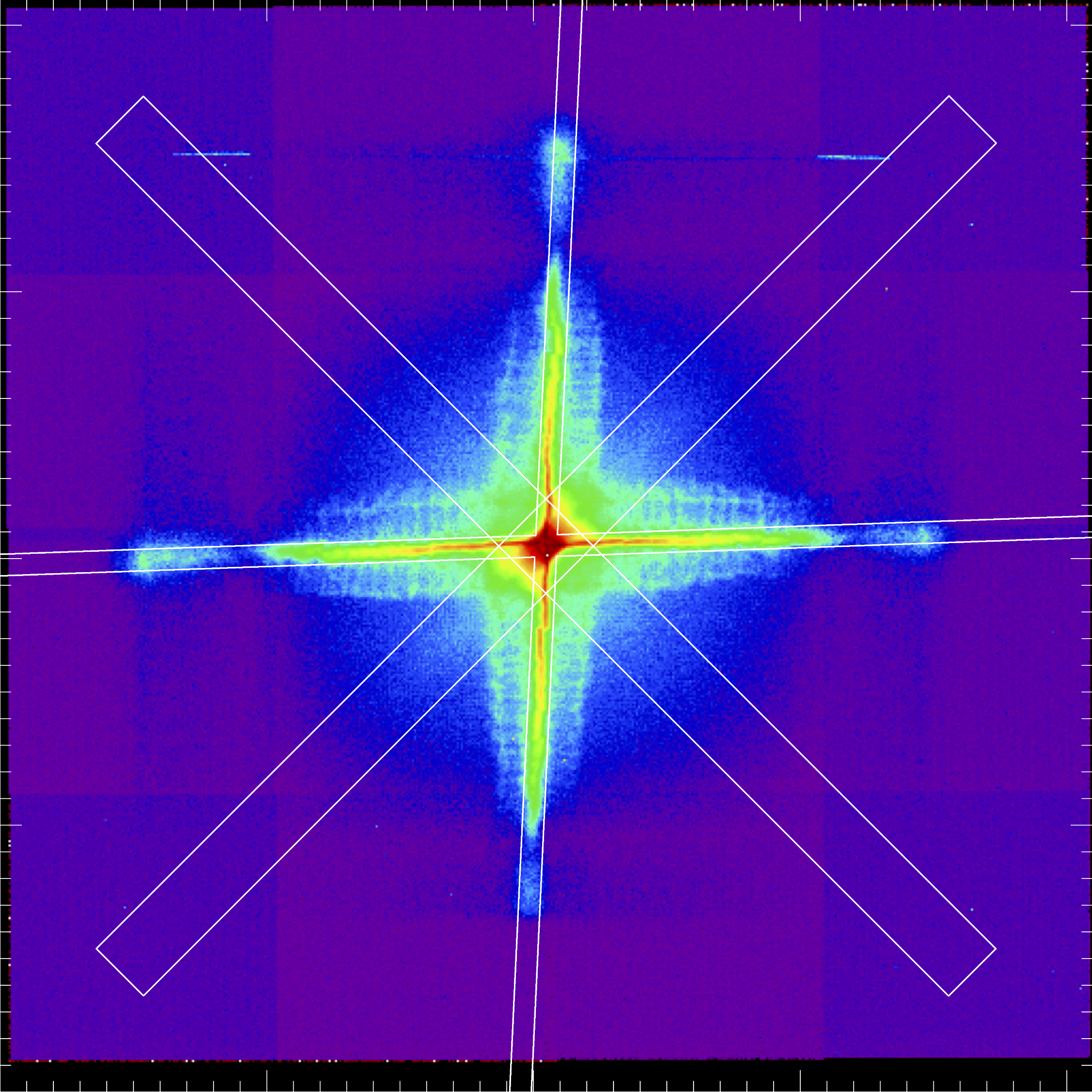}
\includegraphics[width=0.5\textwidth]{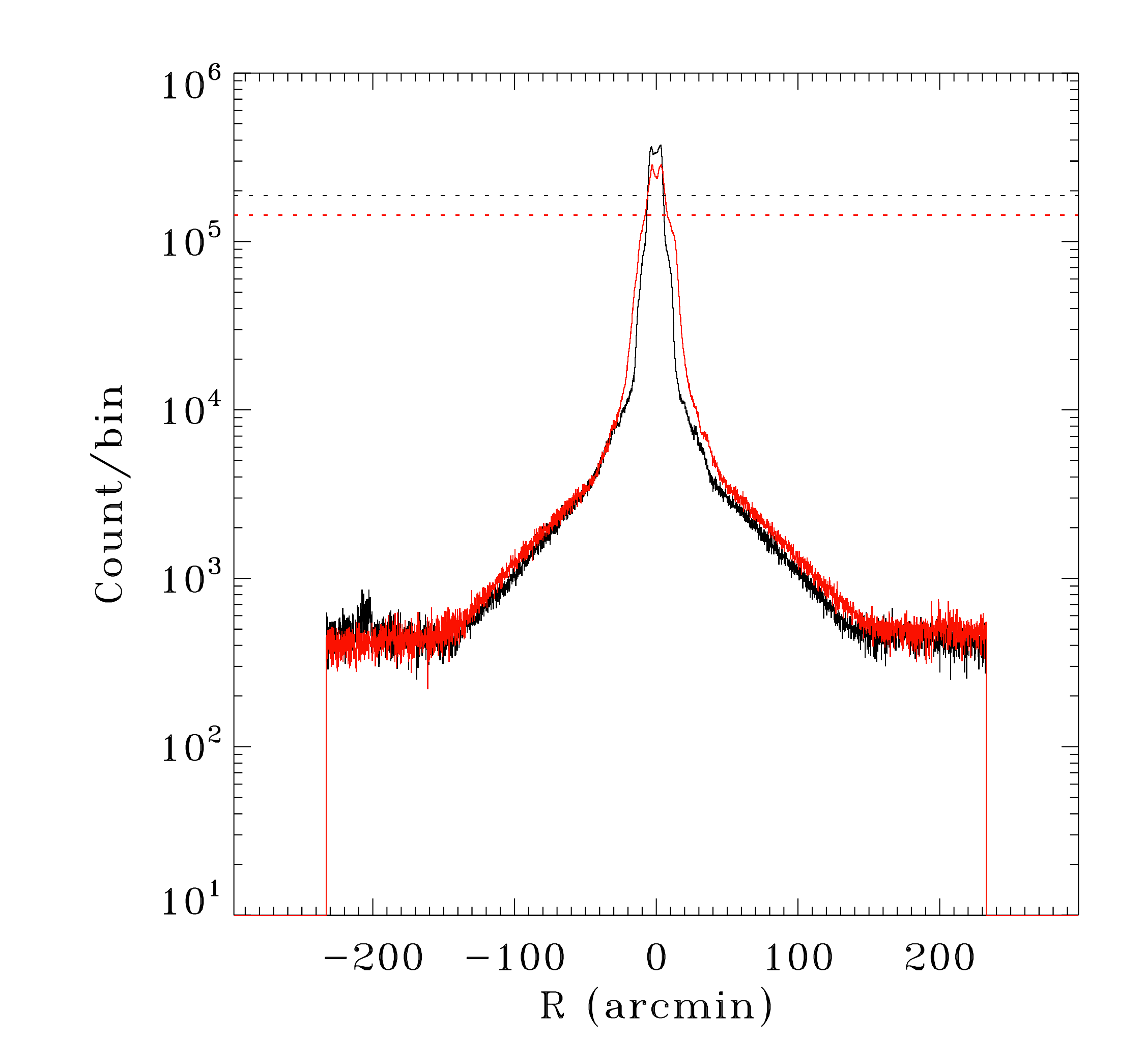}
\caption{
A typical PSF for a micropore optic facet in a logarithmic stretch (left). A profile derived from the diagonal boxes (right). The core is $7.8^\prime$ by $5.6^\prime$. The skewness of the PSF is due to stress from an inappropriate temporary facet holder.
}
\label{fig:psf}
\end{figure*}

A critical single-module assumption is the MPO point-spread function. For the XTRA/MOXI trade study we assume an arcminute-class imaging response, with a representative performance scale of order \(6'\) full width at half maximum and source localizations of order \(1'\) for sufficiently bright detections. The PSF is not circularly symmetric (Figure~\ref{fig:psf}): an MPO produces a focused central spot, two orthogonal cross arms from singly reflected photons, and a uniform background due to photons that pass through the detector without reflection. The shape of the PSF requires source detection based on one-dimensional peak finding in both fundamental directions. For a single MPO rotation, a source of interest can be hidden under the arm of a nearby bright source.

\begin{figure}[H]
\centering
\includegraphics[width=1\textwidth]{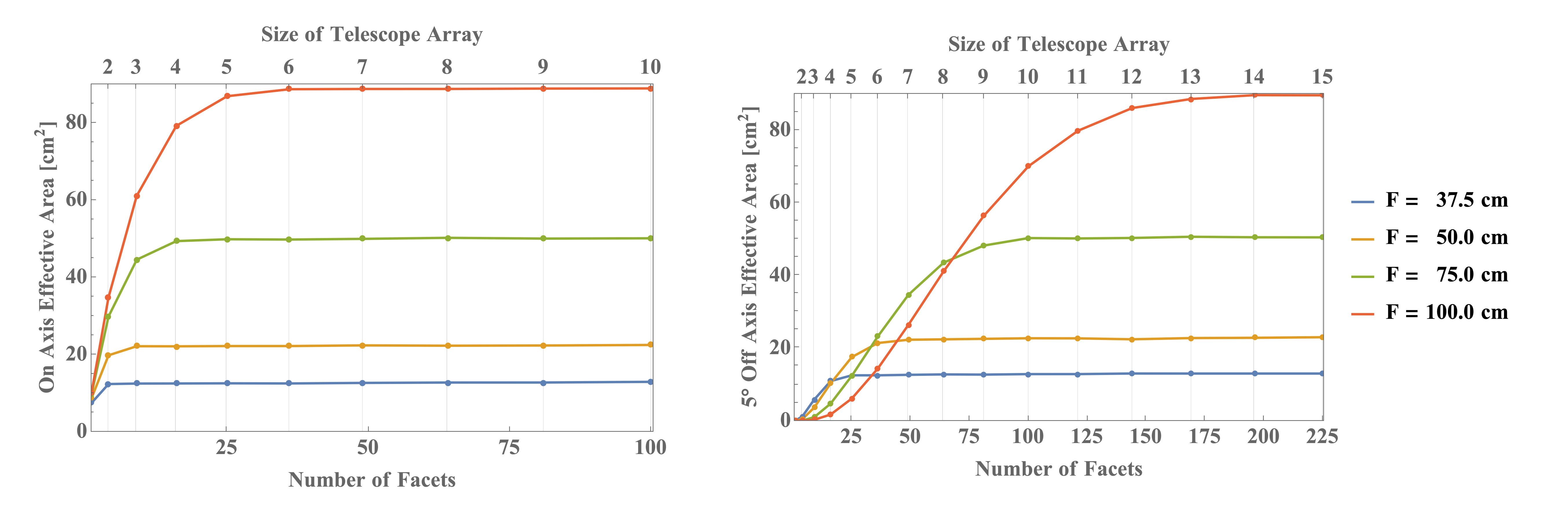}
\caption{
On-boresight (left) and 5$^\circ$ off-boresight (right) effective area at 1~keV as a function of the number of MPO facets for representative focal lengths of 37.5, 50, 75, and 100~cm (bottom scale is total number of MPO facets, top is number on each side for a square arrangement). In all cases the pore depth to width ratio, $L/D$ is 62.5. For a given focal length, the effective area increases initially with the number of facets but saturates as facets at larger off-axis angles contribute less efficiently due to reduced grazing-incidence reflectivity. Beyond this saturation point, additional MPOs primarily increase the FoV rather than the on-boresight effective area. This behavior sets a natural scale for the optimal module size and motivates the use of multiple smaller modules instead of a single large optic at longer focal length.
}
\label{fig:effarea_facets}
\end{figure}

Figure~\ref{fig:effarea_facets}-left shows the predicted on-boresight effective area at 1 keV as a function of the number of MPO facets for representative focal lengths of 37.5, 50, 75, and 100 cm. For a given look direction, $\vec{L}$, only a small number of MPO facets contributes to the effective area since the pore length to width ratio sets the maximum angle from $\vec{L}$, $\Gamma_1$, at which a pore may contribute to the focus of the source at $\vec{L}$; pixels further away would require too many reflections to contribute to the source at $\vec{L}$, but still contribute to the focus of sources in other directions. The maximum area of the optic which can participate in the focus of a source is $\propto(\Gamma_1 F)^2$, where $F$ is the focal length. As a result, making an optic larger than $2\Gamma_1F$ increases the FOV but does not increase the effective area; one can increase the effective area only by increasing $F$, which will decrease the FOV. It also follows that one can replace a monolithic optic with a focal length $F$ with $n$ smaller co-aligned modules with focal length $F/\sqrt{n}$.\footnote{The mathematics of MPOs is more fully explicated in https://pages.jh.edu/kkuntz1/vade\_mecum.pdf.} The use of multiple smaller telescopes does not affect the focal plane detector(s) performance, as the detector physical area for a given solid angle is proportional to $F^2$, hence the total instrument noise per solid angle stays the same. 
For example, a single telescope with 1~m focal length and an array of 10x10 MPO facets has the same effective area, field of view, and focal plane detector noise (and same number of MPOs) as 4 telescopes with 0.5~m focal length and each with an array of 5x5 MPO facets. Smaller optics, correctly aligned to one another, have the advantage of redundancy, better manufacturability, smaller detectors, and greater descope flexibility. 
This equivalence suggests that the overall system design is not uniquely determined by optical considerations. Instead, the choice of focal length and module size should be driven by higher-level system constraints, including focal-plane detector size and format, mechanical envelope, mass budget, redundancy requirements, and manufacturing considerations. In particular, shorter focal length systems offer advantages in compactness and detector matching, while modular configurations provide additional flexibility in survey, pointing, and fabrication strategies.

These considerations led to the LXT and SIBEX configurations, with a 4.5x4.5 cm focal plane detector coupled to a 2$\times$2 MPO module and $F=50$~cm. However, for a SMEX mission, mass and overall system design are also driving parameters, and for XTRA-MOXI such considerations lead to a telescope design with fewer telescopes with a larger FoV of roughly 10 deg x 10 deg. Figure~~\ref{fig:effarea_facets}-right shows the same plot as on the left, but for a source 5~deg from the axis, near the edge of the FoV, where vignetting becomes important. Maintaining a uniform, large effective area over all the FoV lead to an optimized design with 50~cm focal length, 5x5 MPO facets, and a focal plane CCD approximately 9~cm x 9~cm  (see next section for more details). 

\begin{figure}[H]
\centering
\includegraphics[width=0.8\linewidth]{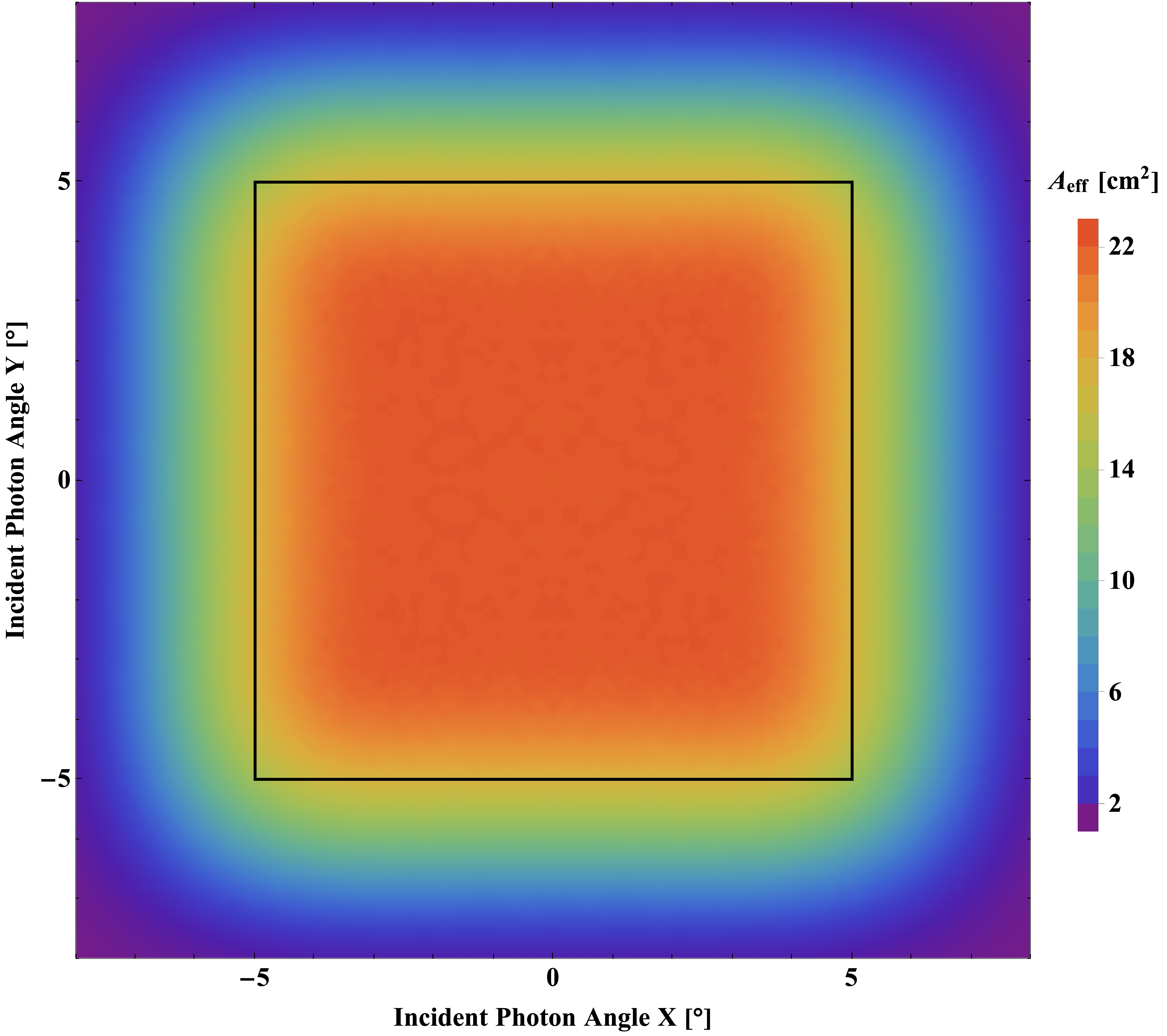}
\caption{
The effective area at 1~keV across the FoV for a single representative MPO module. The response remains uniformly high over a very large FoV, with reduction toward the edges due to vignetting. The square box represents the FoV of  the focal plane detector. The stretch here emphasizes the strongly vignetted region outside the FoV. Comparison with Figure~\ref{fig:vignetting} shows that there is still some energy dependent vignetting within the FoV.}
\label{fig:fov_map}
\end{figure}

\begin{figure}[t]
\centering
\includegraphics[width=0.9\linewidth]{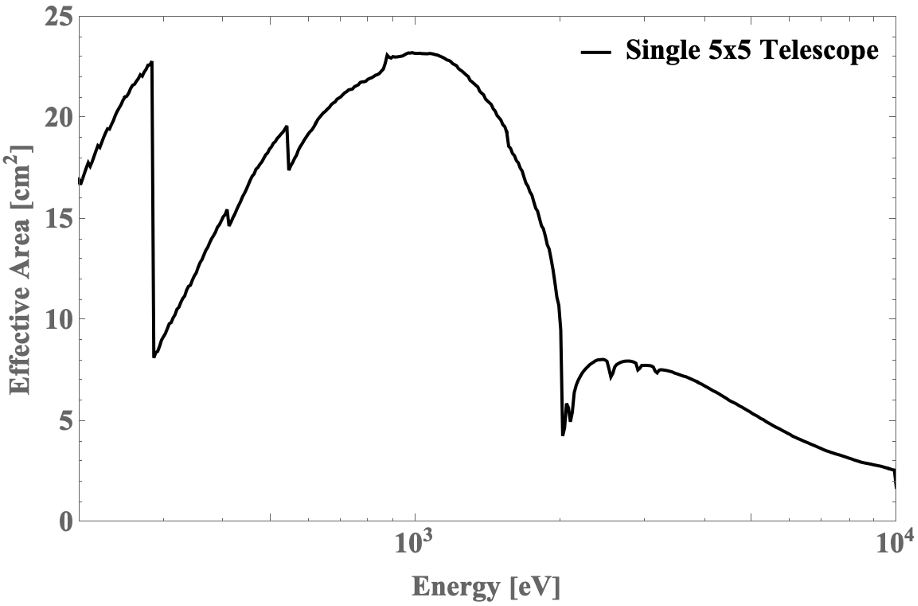}
\caption{
boresight-centered effective area as a function of photon energy for a 5x5 MPO array with 50 cm focal length. The response includes the effect of UV and visible filter and reflects the energy-dependent grazing-incidence reflectivity of the coating, with peak sensitivity in the soft X-ray band and a rapid decline at higher energies where reflectivity decreases. This energy dependence defines the usable bandpass of the instrument and influences the optimization of focal length and pore geometry. 
}
\label{fig:effarea_energy}
\end{figure}

\begin{figure}[t]
\centering
\includegraphics[width=0.9\linewidth]{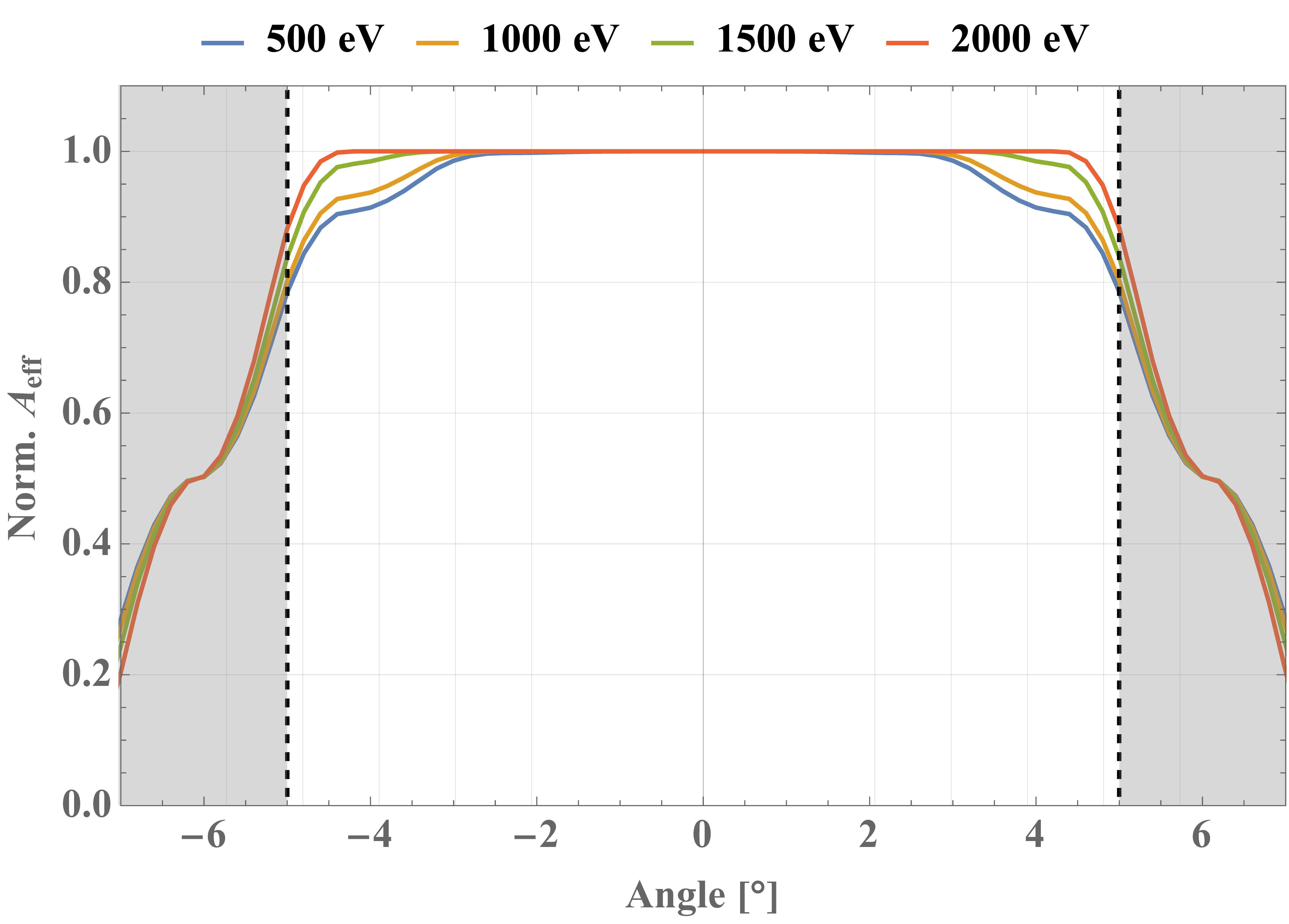}
\caption{
The vignetting function as a function of off-boresight angle for different photon energies. The angular response exhibits an approximately flat large central region followed by an energy-dependent roll-off at larger off-boresight angles. The effective area stays above 80\% of the on-boresight value over the whole MOXI FoV (marked by the dashed vertical lines).}
\label{fig:vignetting}
\end{figure}

The performance characteristics of a representative MPO module are illustrated in Figures~\ref{fig:fov_map}-\ref{fig:vignetting}, which show the spatial distribution of effective area, the boresight-centered effective area as a function of energy, and the vignetting function. Figure~\ref{fig:fov_map} presents the effective area at 1~keV across the FoV. The effective area remains relatively uniform over a large FoV. We have chosen a focal plane detector, shown schematically by the square, that is slightly smaller than half the linear dimension of the optic to ensure that the corners of the detector are only slightly vignetted. This choice also reduces the defocussing due to a curved focal surface. This square module sets the constraints for arranging multiple modules efficiently as discussed in later sections.

Figure~\ref{fig:effarea_energy} shows the on-boresight effective area as a function of photon energy for an iridium-coated MPO. The response is characteristic of grazing-incidence optics, with peak sensitivity in the soft X-ray band and a rapid decline at higher energies set by the rapid decrease of the critical angle for reflection with energy. The low-energy end of the bandpass is determined by the detector response and filter transmission; MOXI is baselining an optical and UV filter identical to those used for the XRISM mission \cite{XRISM} deposited directly on the MPOs as was done for DXL/STORM and LEXI \cite{Collier15,Walsh24}. This behavior defines the operational energy range of the instrument and constrains the optimization of focal length and coating selection.

Figure~\ref{fig:vignetting} illustrates the vignetting function as a function of off-boresight angle for several photon energies. The effective area remains flat within a large central region and stays above 80\% of the on-boresight value over the whole MOXI FoV (marked by the dashed lines). Both energy and off-boresight angle must be considered when optimizing the overall instrument design and survey strategy.

\section{Modular vs. Monolithic Design}

\begin{figure}[H]
\centering
\includegraphics[width=0.8\linewidth]{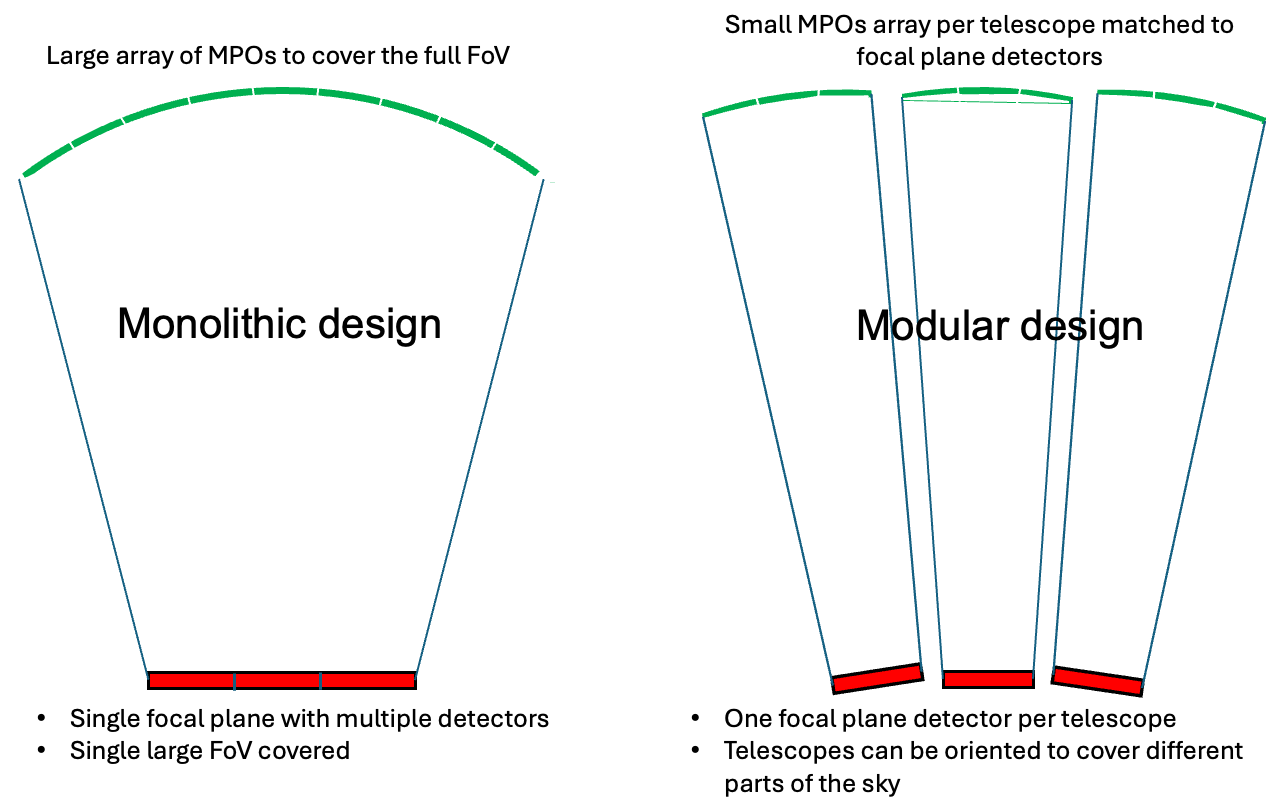}
\caption{
Conceptual comparison of monolithic and modular MPO telescope architectures. In a monolithic design (left), a large optical assembly feeds a contiguous tiled focal plane composed of multiple detectors covering a large FoV with a single instrument, increasing complexity in alignment and calibration. In a modular design (right), multiple smaller telescopes are each paired with a single detector, simplifying the focal plane and enabling scalable, redundant configurations. This comparison highlights the trade-offs between mass, complexity, flexibility, and resilience that drive the system-level design.
}
\label{fig:mod_vs_mono}
\end{figure}

The single-module performance considerations described above naturally lead to two competing architectural approaches for wide-field MPO instruments: a modular design consisting of multiple independent telescopes, and a monolithic design in which a single large optic is coupled to a correspondingly large focal plane (Figure~\ref{fig:mod_vs_mono}).
In a monolithic configuration, a large number of MPO facets are assembled into a single optical structure feeding either a large-area detector or a mosaic of closely packed detectors, and curvature becomes more problematic. This approach minimizes structural duplication and can, in principle, reduce overall mass by eliminating redundant support systems. 
However, monolithic designs introduce significant challenges. The focal plane becomes complex, requiring multiple detectors to be butted together with tight tolerances. Assembly, calibration, and alignment of a large optic are more demanding, and the system is inherently less resilient: a failure or defect in the optical or detector chain can compromise a substantial fraction of the instrument’s capability.

In contrast, a modular approach divides the instrument into multiple smaller telescopes which offers several important advantages. 
The focal plane design for each module is simplified, butting detectors can be reduced or eliminated, and focal plane curvature issues are reduced.
The system can be scaled by replicating a single well-characterized unit. Manufacturing and calibration are more streamlined, as each module can be fabricated, tested, and validated independently. One module can be calibrated while the previous one is integrated and the next one is being fabricated. Failures during manufacturing or operation affect only a single module rather than the entire instrument.
Modular systems are inherently redundant: the loss or degradation of one module does not eliminate the instrument’s functionality, but instead results in a graceful reduction in performance.
Modularity also enables greater flexibility in FoV design and survey strategy. Individual telescopes can be co-pointed to maximize sensitivity, offset to maximize sky coverage, or arranged in more complex geometries—such as the tri-petal configuration described in this work—to optimize both survey uniformity and pointed sensitivity.

The present study for a SMEX-class mission converged on an optimal design that lies between a fully monolithic system, which is lighter but complex and less resilient, and a highly modular system like SIBEX (24 modules) which is flexible but incurs additional mass and system overhead. We adopt an intermediate solution consisting of three identical, independently configurable telescope modules, each composed of a \(5\times5\) array of MPO facets, which can deliver the required performance for both survey {\it and} pointed observations while maintaining manageable complexity, cost, and risk. The tri-petal survey architecture is not an obvious solution; we motivate it in the following sections.

\section{Square-format optics and the need for differential roll}

The square module footprint naturally occurring in MPO telescopes is a main design driver. For a circular field, roll angle is irrelevant, while for a square field, the roll angle provides an extra degree of freedom, controlling where edges, corners, and overlap zones fall.  A no-rotation geometry, in which all three square modules share the same orientation and are merely translated, does not exploit this degree of freedom.  It can produce acceptable instantaneous area, but when tiled over thousands of square degrees it tends to create row-dependent gaps, redundant striping, and non-isotropic overlap patterns.

\begin{figure}[H]
\centering
\includegraphics[width=0.83\textwidth]{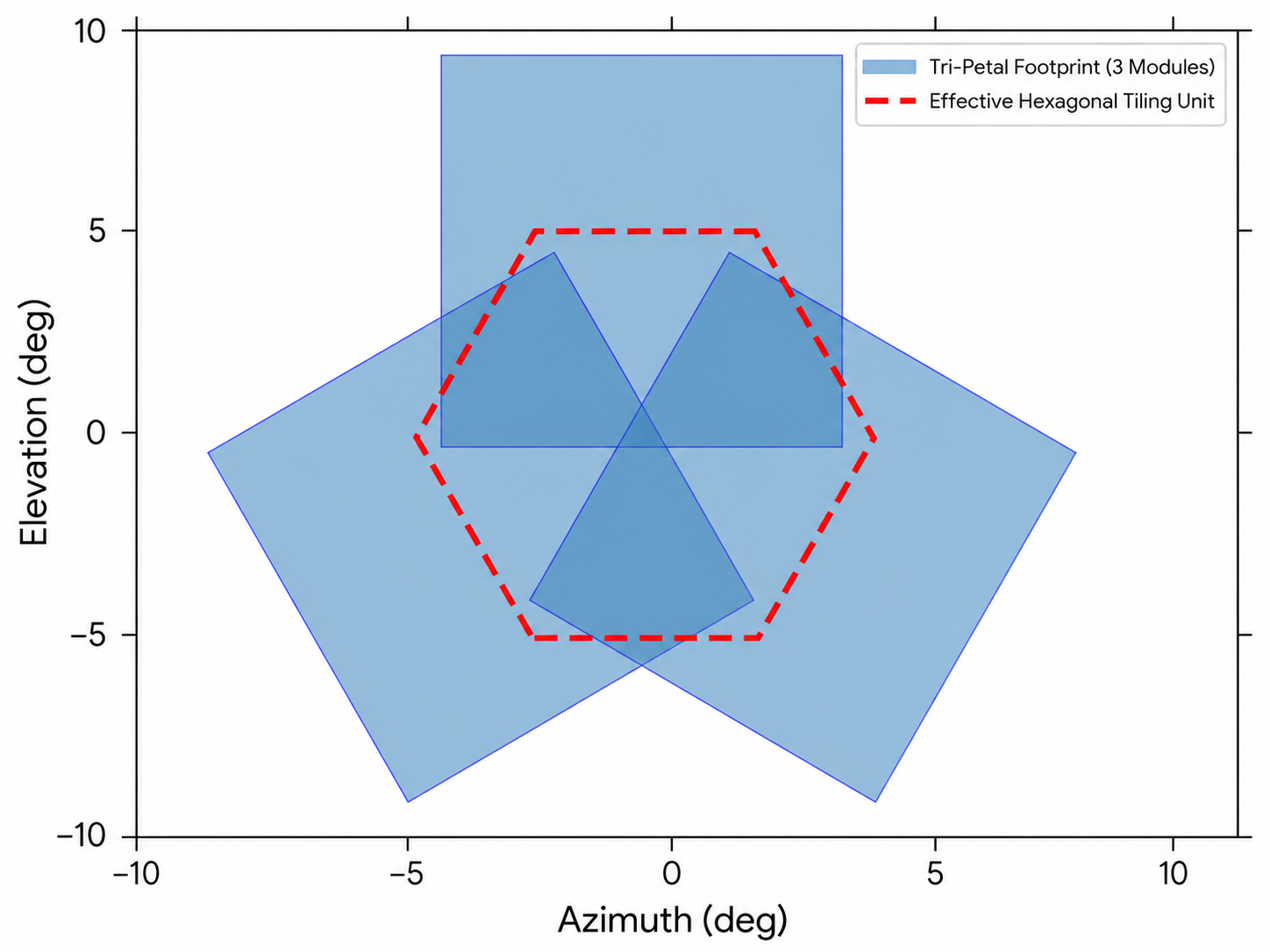}
\caption{Geometric approximation of the three-module tri-petal footprint.  The blue regions show three square-format module fields; the red dashed outline marks the effective hexagonal tiling unit. The differential-roll geometry allows square modules to fill a hexagonal envelope that can be tiled efficiently, even on a sphere. The lightest blue denotes the single telescope coverage or the petal region, the intermediate blue denotes the two telescope overlap or the {\it bridge} region and the darkest blue is the {\it central core} where all three telescopes overlap}
\label{fig:hexagonal_unit}
\end{figure}

An optimized tri-petal arrangement uses both boresight offsets and differential roll.  The three square fields are placed with approximate azimuthal symmetry: their boresights are offset from the common instrument axis, and the square footprints are rolled relative to one another on the sky.  In the XTRA trade-study implementation, three \(10^\circ\times10^\circ\) fields are offset by approximately \(4.6^\circ\) from the instrument axis, with the module rolls chosen so that the square edges and corners interlock rather than forming a common-roll Cartesian pattern. The offset was determined by having an overlap area of order one square degree for pointed observations. This was set to avoid the central overlap getting too close to the edge of the FoV where vignetting reduces the effective area. The resulting compound footprint has a useful envelope that is well approximated by a hexagonal unit cell with the same width as each square.  This effective hexagon, rather than the rectangular bounding box of the three squares, is the natural tiling primitive for the survey. Figure~\ref{fig:hexagonal_unit} illustrates this distinction: the tri-petal geometry is produced by the combination of boresight offsets and relative roll, not by translation alone. For the adopted \(4.6^\circ\) offset, the narrowest scale of the central
overlap is approximately \(0.4^\circ\), or \(24'\), several times larger than
the representative MPO PSF core.  This keeps a target placed in the pointed
core well inside the useful high-response region of all three modules; larger
offsets would increase geometric survey area but erode the pointed core, while
smaller offsets would increase the core at the cost of wide-field grasp.

This geometry is also attractive from a spacecraft-accommodation standpoint. The three telescope units remain compact, replicated modules placed approximately symmetrically about a common instrument axis, rather than forming a single large monolithic optical assembly. This balanced arrangement naturally allows a second instrument to be placed in the center of the modules and aligned with the highest overlap regions. The differential roll is therefore not a mechanical complication imposed on an otherwise simple payload; it is a field-geometry optimization applied to a modular layout that is already well matched to spacecraft accommodation.

The design can be summarized by three regions. The ``outer petals'' are covered by one module and provide the baseline survey sensitivity.  The ``bridges'' are covered by two modules and provide an intermediate-depth region. The central core is covered by all three modules and provides the natural aim point for pointed observations. Since the bridges are formed by the edges of the detectors, the CCDs can be oriented to have their readout node(s) located on that edge so that the bridges experience lower charge transfer inefficiency (CTI), and thus better energy resolution. Clever partial clocking of the CCDs may allow them to operate in a high speed mode for high time resolution spectroscopy of bright sources. Table~\ref{tab:tripetal_depth_regimes} summarizes the observational properties of the tri-petal design. Note that the tri-petal footprint allows both localization and follow-up of transient alerts with small error boxes and synoptic X-ray surveys of large fields.

Note that the effective PSF in the region of greatest(intermediate) overlap effectively has 12(8) arms, and thus more closely approximates a ``traditional'' X-ray PSF produced by Wolter type optics, and may allow more traditional methods of point source detection.

\begin{table}[ht]
\centering
\caption{Representative area and relative-depth regimes for a single XTRA-like
tri-petal pointing. Areas are approximate geometric values for three
differentially rolled \(10^\circ\times10^\circ\) square modules forming a
\(\sim241\ \degsq\) compound field.}
\label{tab:tripetal_depth_regimes}
\small
\renewcommand{\arraystretch}{1.2}
\begin{tabular}{lcccc}
\hline
\textbf{Region} & \textbf{Module depth} & \textbf{Approx. area} & \multicolumn{2}{c}{\textbf{Flux-limit improvement factor$^*$}} \\
\cline{4-5}
                &                       &                       & \textbf{Background limited} & \textbf{Photon limited} \\
\hline
Petals & 1 & \(\sim183\ \degsq\) & \(1\) & \(1\) \\
Bridges & 2 & \(\sim56\ \degsq\) & \(\sqrt{2}\) & \(2\) \\
Central core & 3 & \(\sim1.4\ \degsq\) & \(\sqrt{3}\) & \(3\) \\
\hline
\end{tabular}

\vspace{2pt}
\begin{flushleft}
\footnotesize
* For the representative MOXI sensitivity model used here, the
transition between photon-limited and background-limited behavior occurs at
exposure times of order \(10^3\) s.  The exact break point depends on effective
area, vignetting, background, source spectrum, PSF/detection-cell size,
detector gaps, and the adopted detection algorithm.
\end{flushleft}
\end{table}


\section{Single-pointing geometry: survey field plus pointed core}

The geometric tri-petal footprint is a useful first-order description of the survey, but the final sensitivity map must include the true off-axis response of each lobster-eye module.  In the simplified treatment, each telescope is represented as a square field with a uniform effective area inside the nominal FoV and zero response outside.  This approximation is useful for understanding the petal, bridge, and core geometry, but it overstates the sharpness of the field boundary and ignores the energy-dependent vignetting of the micropore optics. 

For the flight-like sensitivity calculation, the binary square footprint should therefore be replaced by the measured or modeled single-telescope vignetting function,
\begin{equation}
  V(E,\theta_x,\theta_y) =
  \frac{A_{\rm eff}(E,\theta_x,\theta_y)}
       {A_{\rm eff}(E,0,0)} .
\end{equation}
where \(A_{\rm eff}(E,\theta_x,\theta_y)\) is the effective area of one module at photon energy \(E\) and angular offset \((\theta_x,\theta_y)\) from that module's boresight. One advantage of MPO optics is that, due to their geometry $x$ and $y$ are independent and thus $V(E,\theta_x,\theta_y)\simeq A_{\rm eff}(E,\theta_x)/A_{\rm eff}(E,0)\times A_{\rm eff}(E,\theta_y)/A_{\rm eff}(E,0)$.  The exposure-weighted effective area at a sky position \(\boldsymbol{\Omega}\) is then the sum over all contributing modules and visits,
\begin{equation}
  A_{\rm survey}(E,\boldsymbol{\Omega}) =
  \sum_{k=1}^{N_{\rm visit}}
  \sum_{m=1}^{N_{\rm mod}}
  t_k \,
  A_{{\rm eff},m}
  \left[
  E,
  \theta_{x,m,k}(\boldsymbol{\Omega}),
  \theta_{y,m,k}(\boldsymbol{\Omega})
  \right],
\end{equation}
where \(t_k\) is the exposure time for visit \(k\), and \((\theta_{x,m,k},\theta_{y,m,k})\) is the sky position transformed into the local coordinate frame of module \(m\), including the module boresight offset and roll angle.

Note that the effective area calculation includes the focused, partially focused, and unfocused photons. The effect of the unfocused photons must be addressed in the sensitivity calculations.

\begin{figure}[t]
  \centering
  \includegraphics[width=0.85\linewidth]{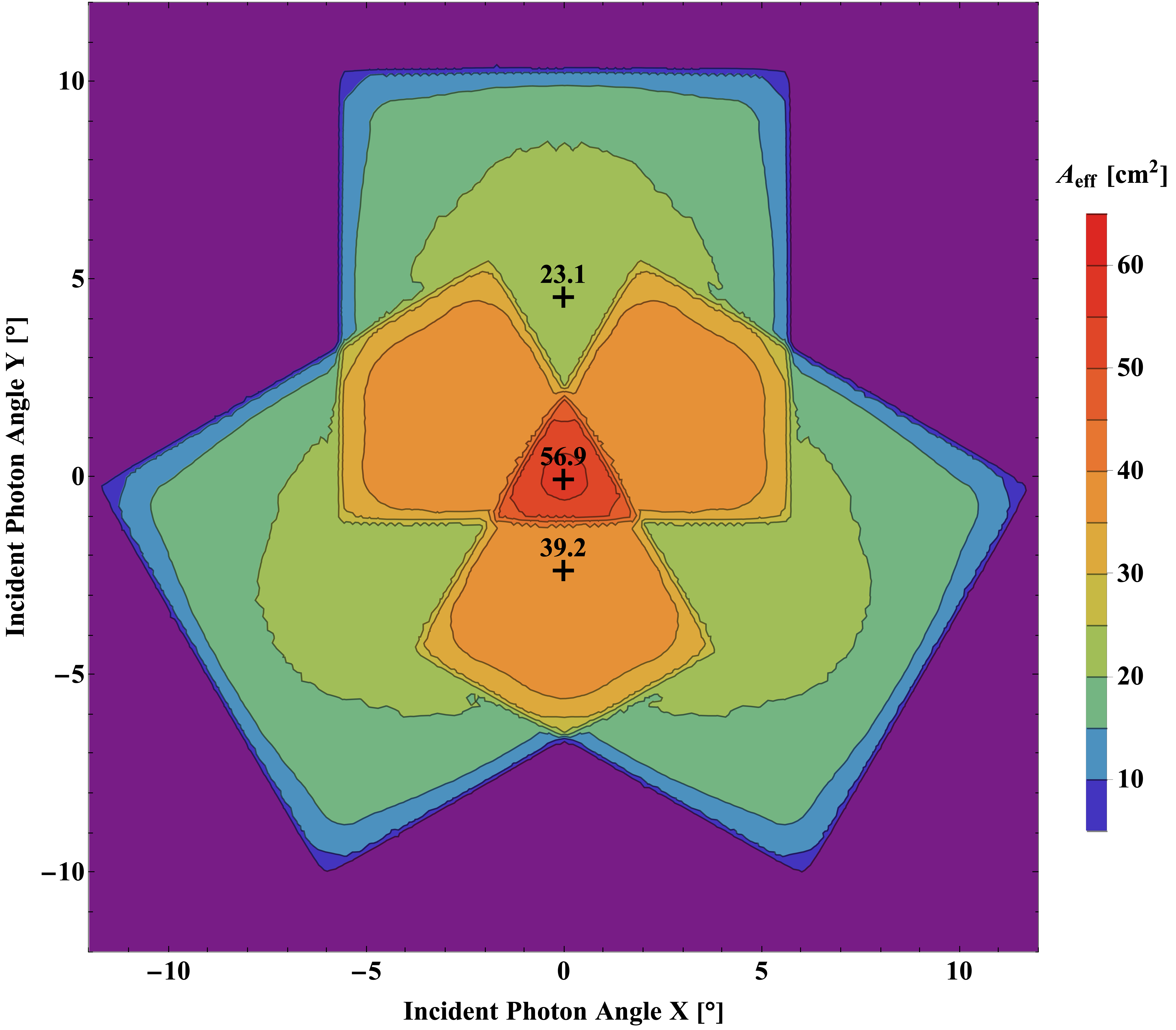}
  \caption{
  Full MOXI effective area at 1~keV. 
  The idealized square-field geometry used to define the tri-petal footprint is
  replaced in the final sensitivity calculation by the energy-dependent effective
  area \(A_{\rm eff}(E,\theta_x,\theta_y)\).  This converts the integer
  one-, two-, and three-module overlap map into a smoothly varying
  vignetting-weighted exposure map. This does not show the area covered by the focal plane detector.
  }
  \label{fig:moxi_vignetting}
\end{figure}

Figure \ref{fig:moxi_vignetting} shows the resulting combined effective area. The nominal single-module petals, two-module bridges, and three-module core become regions of smoothly varying effective exposure rather than discrete depth zones.  In particular, the bridge regions are not simply ``twice as deep'' everywhere: their sensitivity depends on the product of two off-axis vignetting functions, which can either reinforce the survey depth or fall below the idealized geometric expectation near the corners of the square fields.  The central overlap region remains the natural aim point for pointed observations because all three modules contribute at close to their peak effective area.

\begin{figure}[t]
\centering
\includegraphics[width=0.72\textwidth]{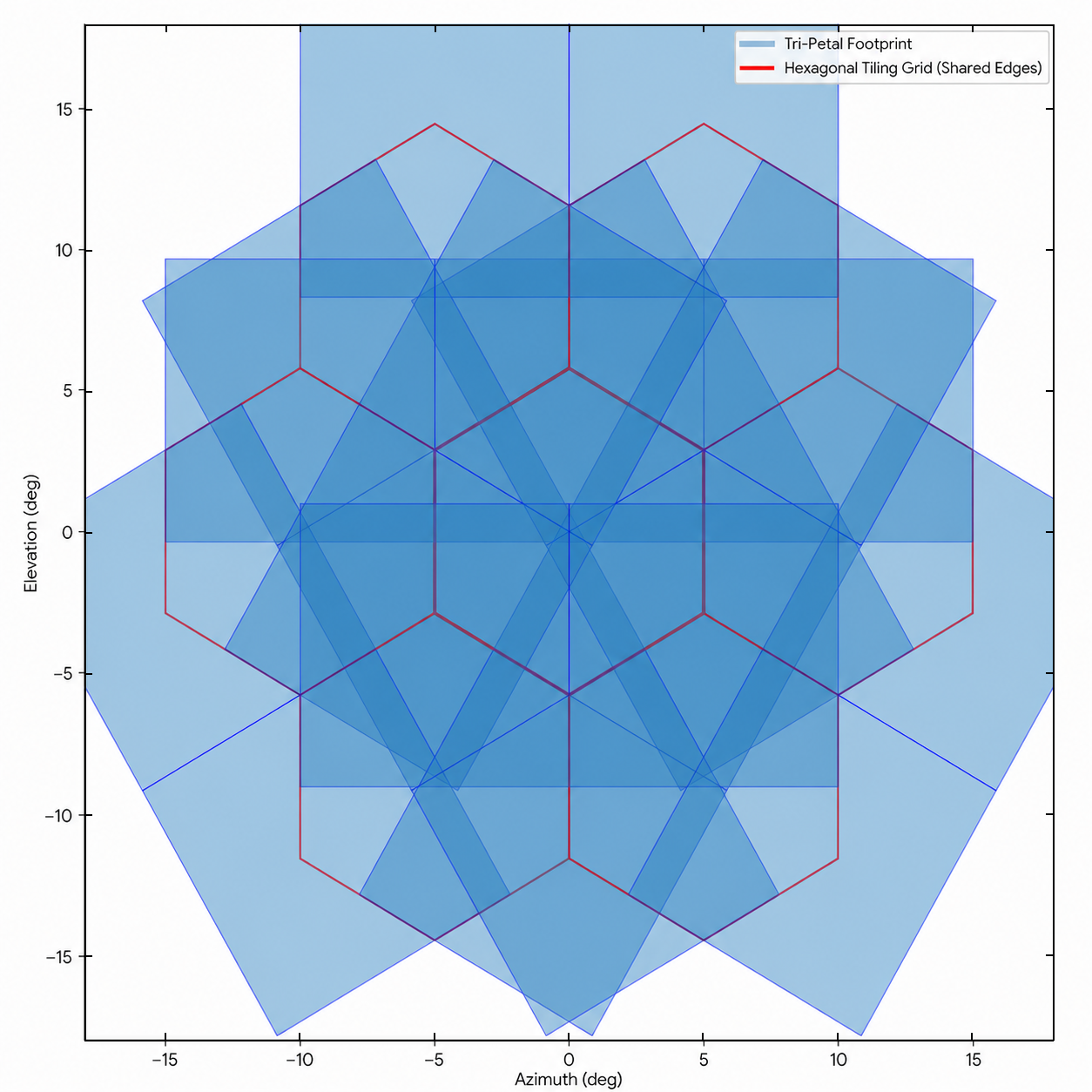}
\caption{Seven-cell illustration of the hexagonal tiling concept.  The full tri-petal footprint is repeated on a center-plus-six-neighbor lattice.  The shared flat edges are the large-area survey advantage of the differentially rolled square-field solution. Note: The hexagon has been rotated by 30 degrees relative to Figures \ref{fig:hexagonal_unit} and  \ref{fig:moxi_vignetting}}
\label{fig:seven_hex}
\end{figure}

\section{Hexagonal-knit survey tiling}

The Swift comparison in the introduction demonstrates why the survey tiling must be treated as a first-order design variable. While Swift with its $23^\prime\times23^\prime$ FoV has produced a valuable serendipitous XRT survey, it would struggle to cover a large ($10^3\degsq$) FoV within a short period of time without optimising strategies based on reasonable assumptions \cite{evans2016}. The Einstein Probe lies near the opposite extreme, providing {\it shallow} coverage of roughly half the sky in three orbits, and depending upon another instrument, the EP/FXT, to provide narrower-field follow-up of selected transients \cite{Yuan2022}. Rather than simply maximizing sky coverage, the tri-petal/Hexagonal-Knit strategy is a prescribed survey mode designed to produce a controlled exposure and revisit pattern of an extended region with the same wide-field focusing instrument used for discovery and follow-up.

This distinction is important for modern time-domain and multimessenger astronomy.  The regions requiring follow-up span from arcsecond-scale optical or X-ray positions, which can be observed efficiently with conventional narrow-field Wolter-I telescopes, to hundreds or thousands of square degrees for gravitational-wave, neutrino, poorly localized gamma-ray, or fast-transient alerts.  The latter case requires not only a large instantaneous field of view, but also a tiling strategy that can cover a large probability region quickly while producing a relatively uniform exposure map.

The Rubin/LSST survey \cite{Zeljko2019RubinSurvey} is an example of a representative optically rich time-domain footprint and alert stream. A single MOXI tri-petal pointing can cover many Rubin fields. Besides following up LSST transient alerts, science cases can be made for routine follow-up of subsets of the LSST footprint, be it the deep-drilling fields, or some fraction of the Rubin/LSST footprint at a particular cadence. The tri-petal geometry and observing strategy is sufficiently flexible to handle all of those cases efficiently.

The survey does not simply stack bounding boxes.  It uses a high-density hexagonal lattice in which adjacent pointings interlock (Fig.~\ref{fig:seven_hex}).  Hexagonal lattices are the natural reference geometry for efficient coverage with circular or approximately hexagonal footprints: in the idealized problem of covering the plane with equal circles, the optimal covering is achieved by a hexagonal lattice \cite{Kershner1939}. In the XTRA case, the differentially rolled square modules produce an effective hexagonal footprint, so the same logic applies: the natural survey lattice is hexagonal rather than rectangular.

\begin{figure}[H]
\centering
\includegraphics[width=0.92\textwidth]{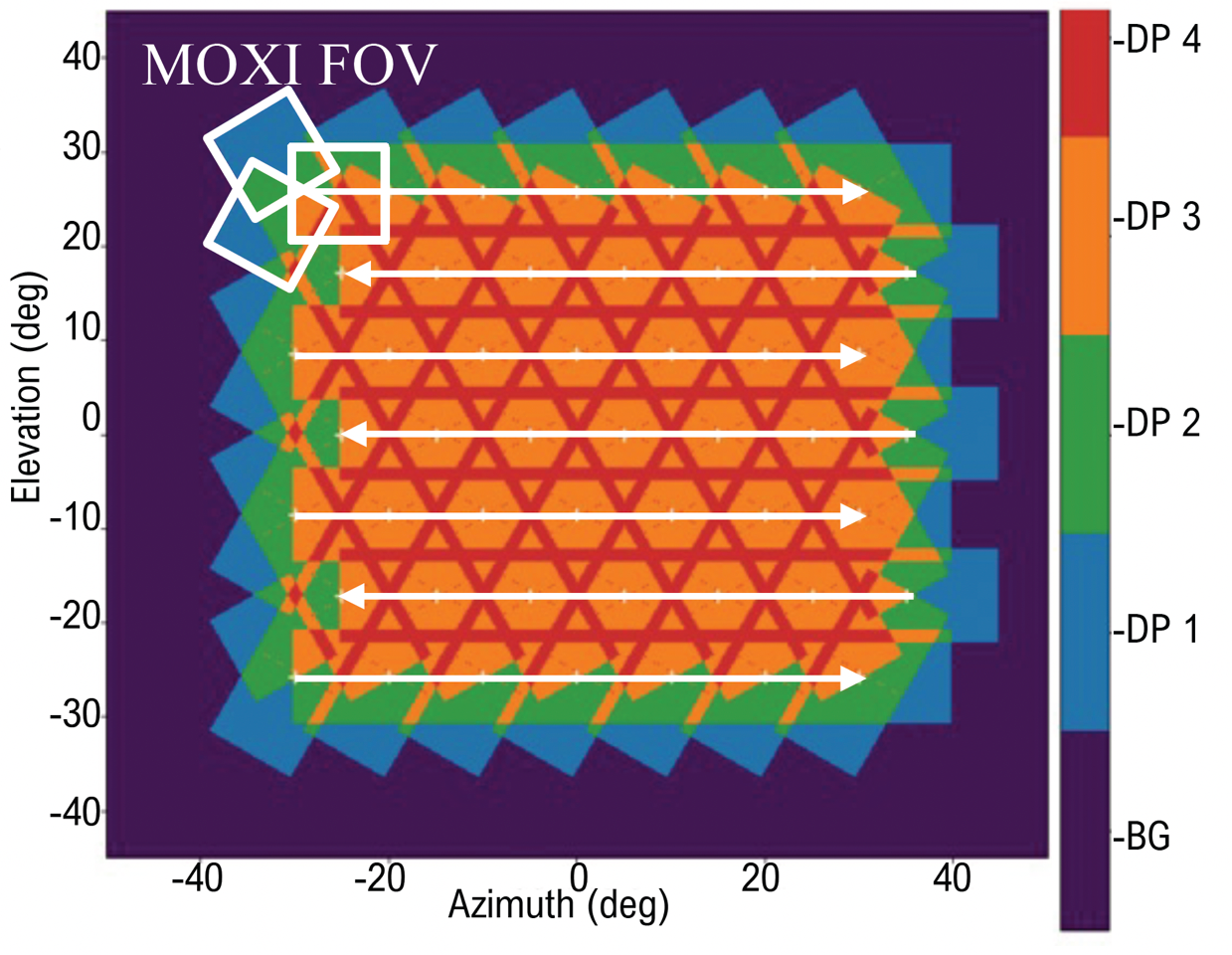}
\caption{Representative ($7\times7$) block of the optimized tri-petal survey simulation.  The white crosses mark pointing centers.  The dense
hexagonal-knit spacing closes gaps and builds a tiered depth map: blue is depth
1, green is depth 2, orange is depth 3, and red is depth 4 or greater.  This
figure shows why the tri-petal geometry should be treated as a survey lattice,
not merely as a single pointing.}
\label{fig:lsst_block}
\end{figure}

We tightened the pointing grid from the sparse configuration
shown in Fig.~\ref{fig:seven_hex} to a final horizontal-by-vertical pitch of
approximately \(10.0^\circ\times8.6^\circ\).  At this spacing, the outer
regions of adjacent tri-petal pointings overlap: areas that would otherwise be
gaps instead become additional exposure.  The resulting survey is not a set of
isolated fields, but a continuous high-depth region with an
embedded lattice of still greater depth (Figure~\ref{fig:lsst_block}).

In the central region of the representative \(7\times7\) simulation, roughly
59\% of the area is at depth 3 and roughly 41\% is at depth 4 or greater,
corresponding to an average depth of about 3.5 visits per sky position.  In the
optimized lattice the representative four- and five-view regions provide
idealized background-limited flux-limit improvement factors of \(2\) and
\(\sqrt{5}\), respectively.

For nearby approximately Euclidean populations,
a background-limited flux-limit improvement of \(2\)--\(\sqrt{5}\) corresponds
to an ideal accessible-volume increase of approximately \(3\)--\(4\).  This is
only a scaling argument: the realized gain depends on the luminosity function,
absorption, transient duration, spectrum, host environment, telemetry
thresholds, onboard classification, and follow-up completeness.  The quoted
depth fractions and visit counts are survey-simulation outputs, not fundamental
constants of the geometry, and can be re-optimized
at the mission-planning level for different exposure times,
sky areas, or allocations to GRB follow-up and deep-drilling fields.

\subsection{Robustness and degraded operation}

\begin{figure}[H]
\centering
\includegraphics[width=0.86\textwidth]{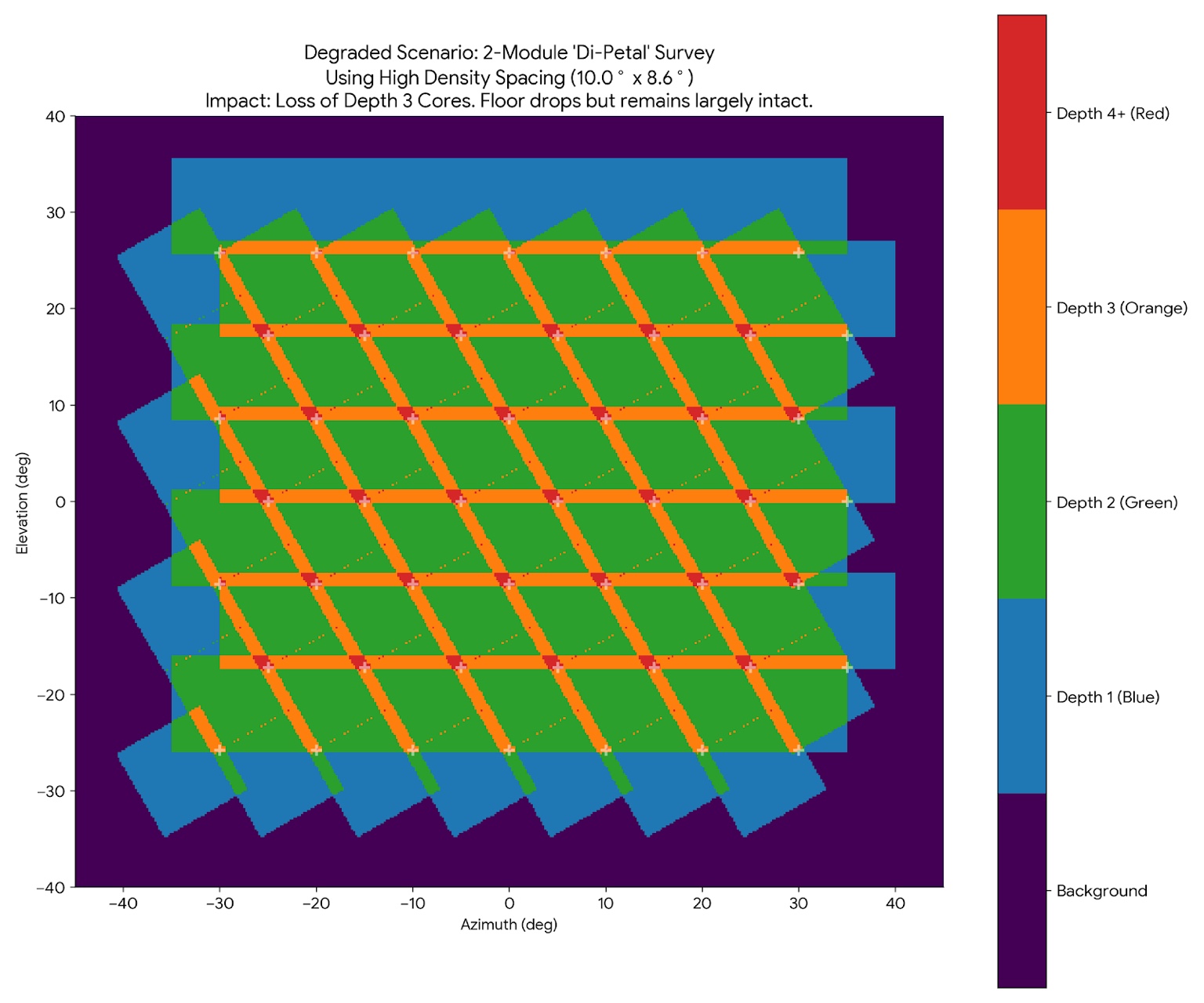}
\caption{Representative degraded two-module case from the trade study.  Loss of one module reduces the single-pointing central region from three-module to two-module overlap, but the high-density survey lattice preserves a connected depth-2 mesh and higher-depth intersections.  This illustrates why the interlocking survey geometry is robust to module loss or descoping.}
\label{fig:degraded}
\end{figure}

A dense tri-petal survey lattice also provides robustness. If one module is lost, the central pointed region is reduced from three-module to two-module overlap, but the interlocked survey lattice does not collapse (Figure \ref{fig:degraded}).  The field becomes a two-module or modified L-like configuration with lower peak depth, but the high-density spacing preserves continuous coverage and retains a depth-2 floor with deeper intersections.  This is a graceful-degradation mode for survey science, although it is not equivalent to the full tri-petal for pointed observations.

\subsection{Temporal sampling}

The survey lattice is most efficiently traversed using a serpentine scan;  adjacent pointings are separated by local steps of order 10$^\circ$, and the scan direction reverses on alternate rows. This sequence avoids long flyback slews, improves observing efficiency, and controls reaction-wheel momentum buildup. The tri-petal geometry also creates multiple time baselines.  For example, a Rubin/LSST-scale survey field of roughly $2\pi$ steradians (18,000 $\degsq$) requires $\sim210$ pointings. With 1 ks exposure per pointing, a full pass requires roughly 3.1 days for an assumed observing efficiency near 86\%. Overlaps between sequential pointings provide rapid revisit times near 20 minutes.  Vertical and diagonal overlaps provide intermediate revisit times from tens of minutes to several hours, depending on row position.

This observation structure is important for transient science because the survey is not only a static depth map.  It is also a temporal sampling pattern.  For fast X-ray transients, shock breakouts, and early GRB afterglow behavior, the 20 minute to few-hour overlap cadence can be scientifically valuable.  For tidal-disruption events, AGN variability, QPEs, and other slower sources, the multi-day global cadence and accumulated depth are more important.

\subsection{Comparison with no-rotation and Cartesian alternatives}

\begin{figure}[H]
\centering
\includegraphics[width=0.43\textwidth]{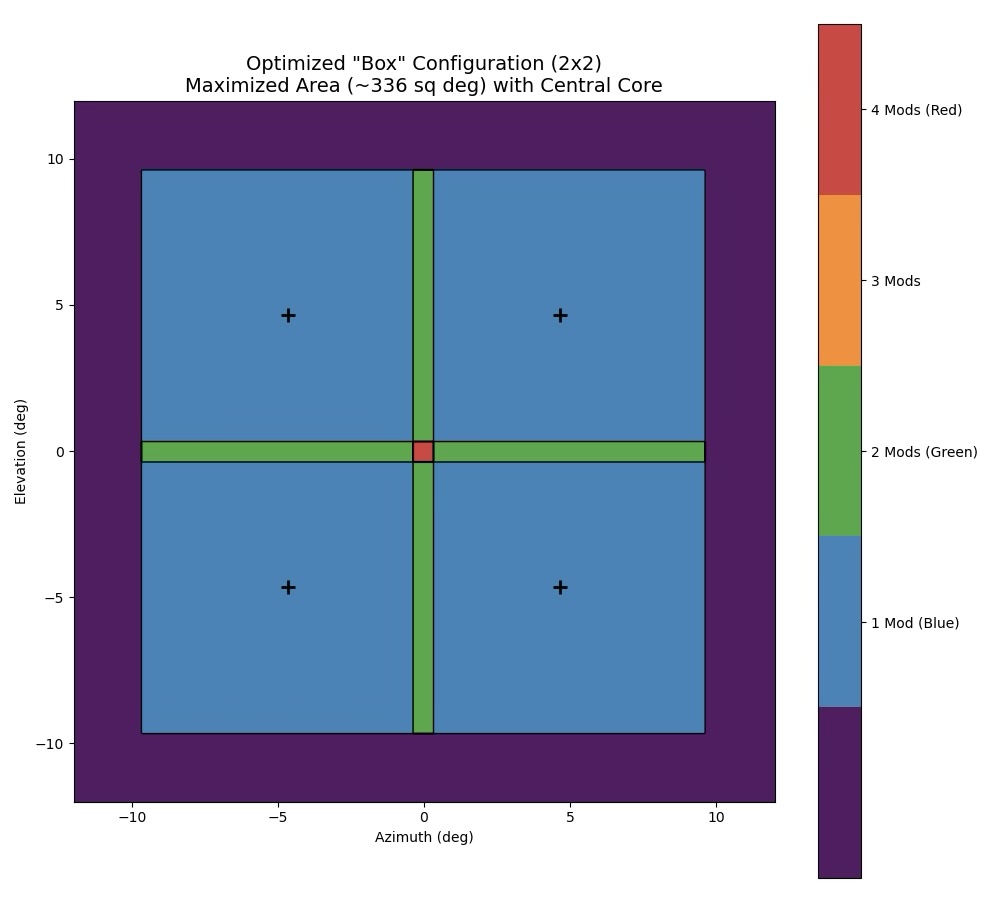}\includegraphics[width=0.43\textwidth]{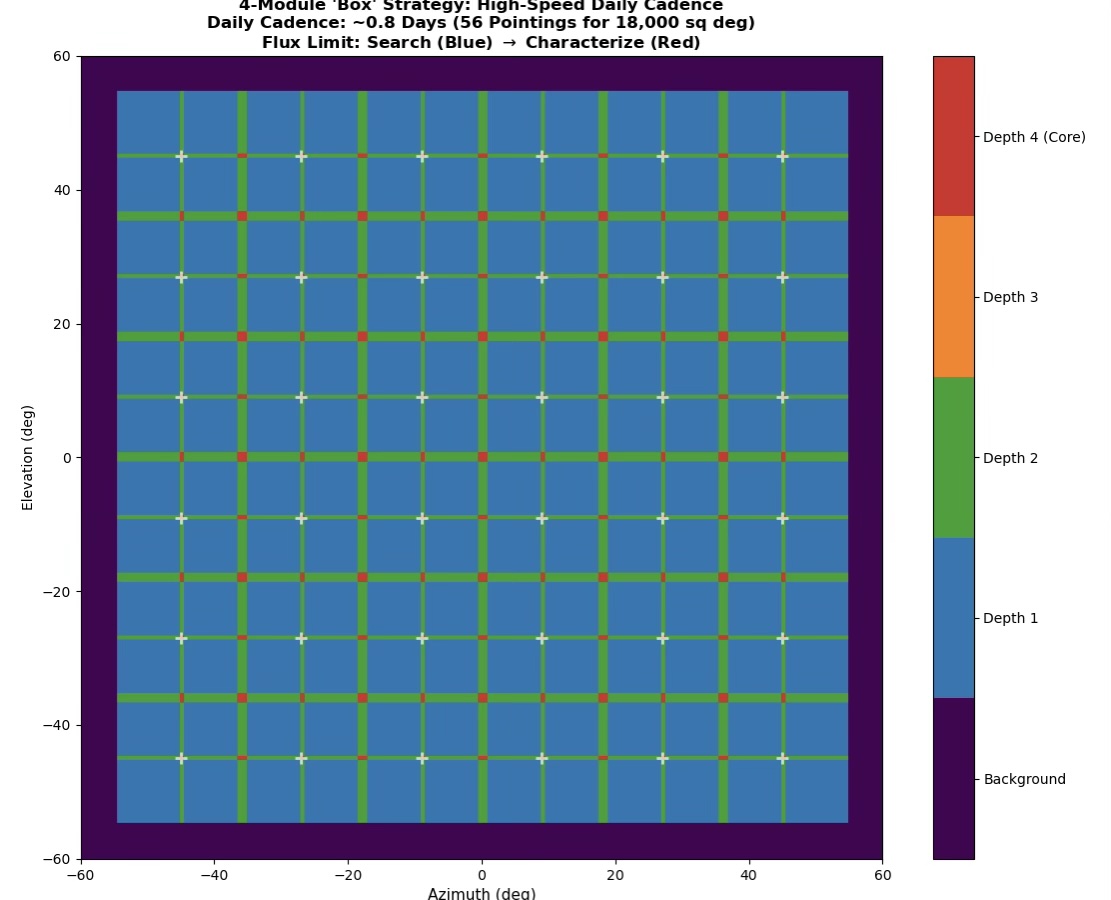}
\includegraphics[width=0.43\textwidth]{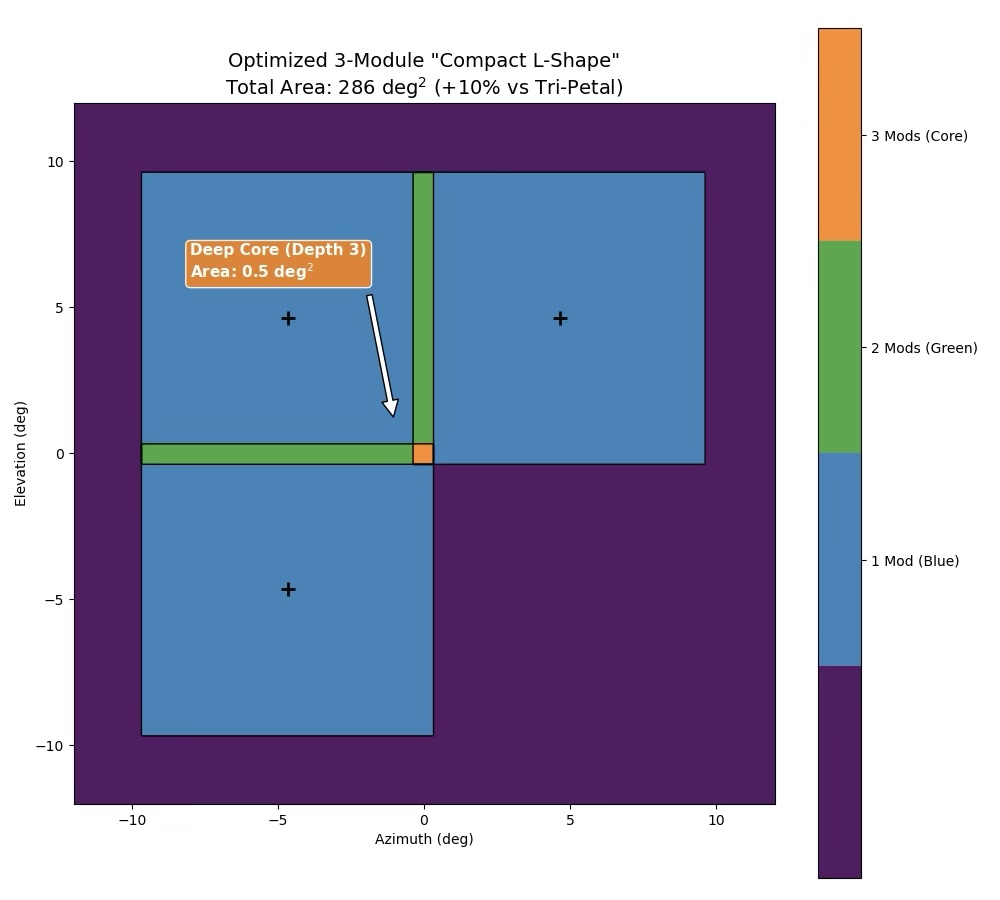}\includegraphics[width=0.43\textwidth]{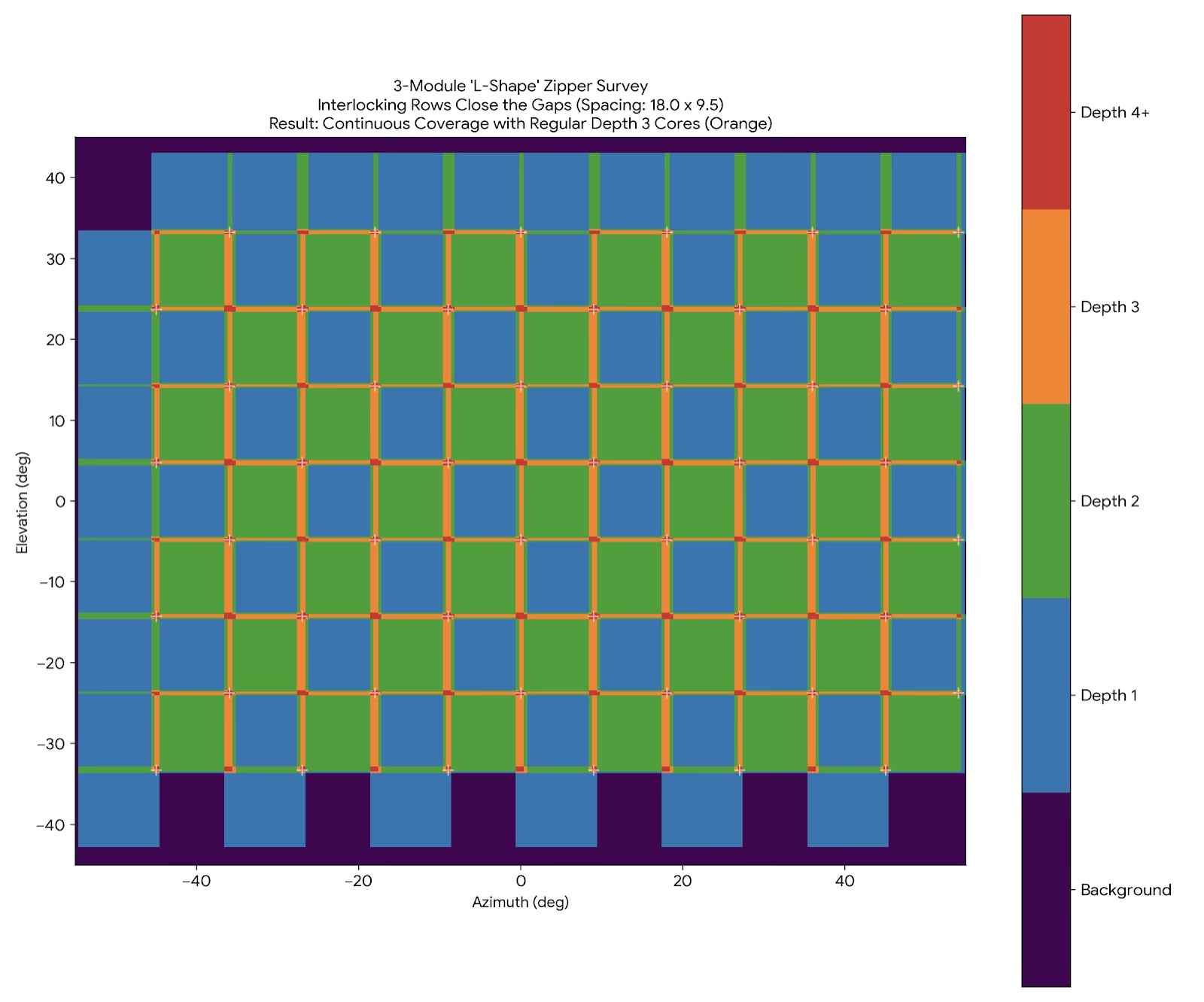}
\caption{A four-module FoV formed with the same criteria used for the tri-petal (top left) and an example \(6\times6\) tiling covering \(336~\mathrm{deg^2}\)(top right). A four-module FoV after loss of one module (bottom left) and an example of tiling such a shape (bottom right). The tiling of an L-shaped FoV requires twice the survey times and produces a strongly structured non-uniform exposure time.}
\label{fig:four}
\end{figure}

We have considered alternatives including four-module boxes, 
compact L-shapes, and three-module T-shapes.  These geometries can be useful for specific goals, but they do not provide the same compromise as the differentially rolled tri-petal.

The four-module box constructed with the same overlap for the core as used for the tri-petal pattern (Figure~\ref{fig:four}) provides $\sim50$\% more instantaneous area, a $\sim33$\% faster survey, $\sim25$\% more short GRBs, and $\sim15$\% more long GRBs, while the tri-petal survey is twice as deep and more uniform. Further, in the regions of overlap for the four-module box the PSF arms are aligned so that faint sources of interest are easily lost under the arms of nearby brighter sources. The four-box module quickly encounters the standard sphere tiling problem with large area mosaics, which the tri-petal does not. Loss of a module on orbit would produce an L-shaped FoV that is difficult to tile efficiently. One needs to  shift every other row horizontally so that the "nose" of the L-shape in one row tucks into the "empty corner" of the L-shape in the row above it.  This strategy requires double the survey time, and produces a strong checkerboard pattern in exposure time (Figure~\ref{fig:four}, lower panels).

\begin{figure}[H]
\centering
\includegraphics[width=0.43\textwidth]{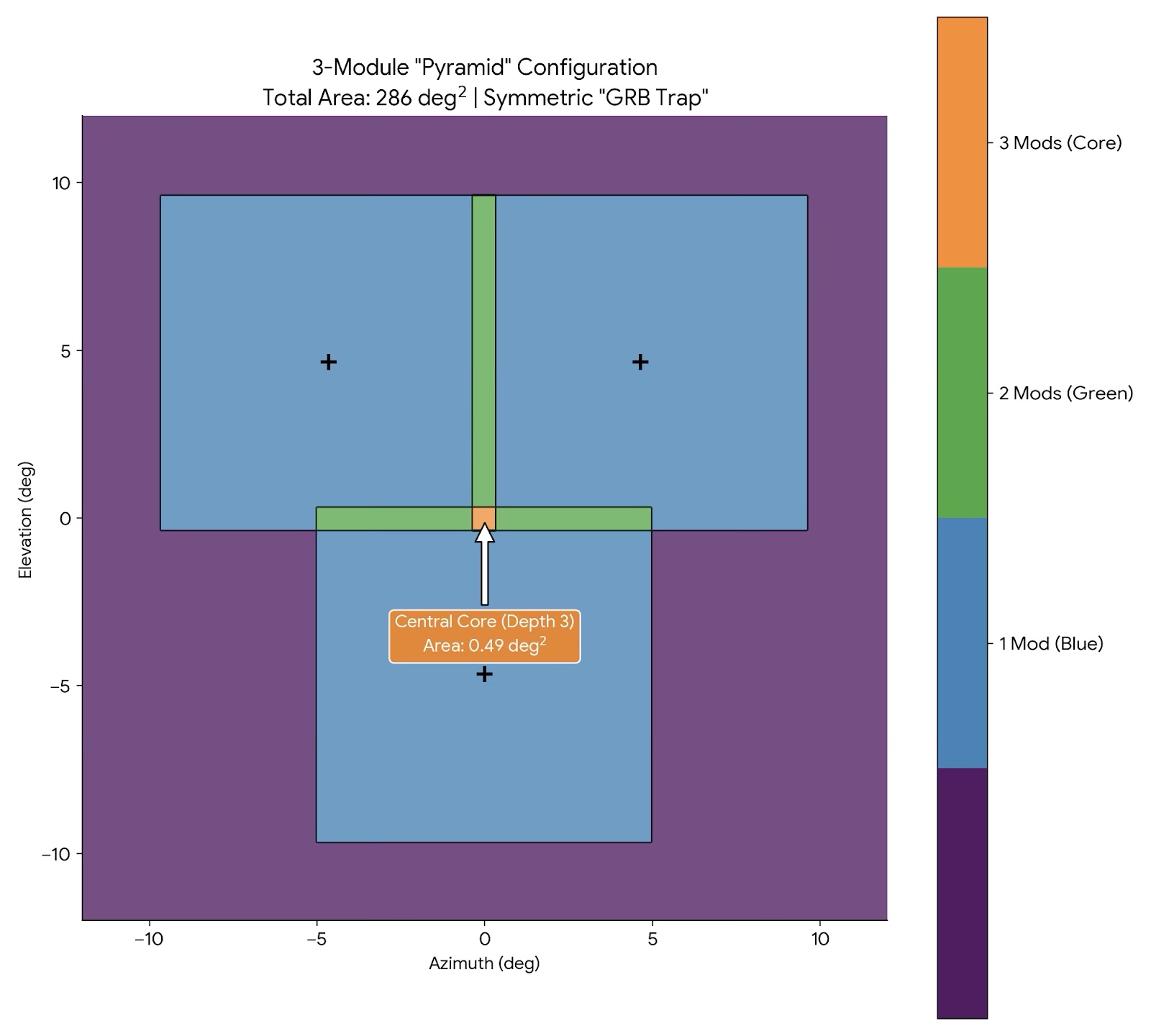}\includegraphics[width=0.43\textwidth]{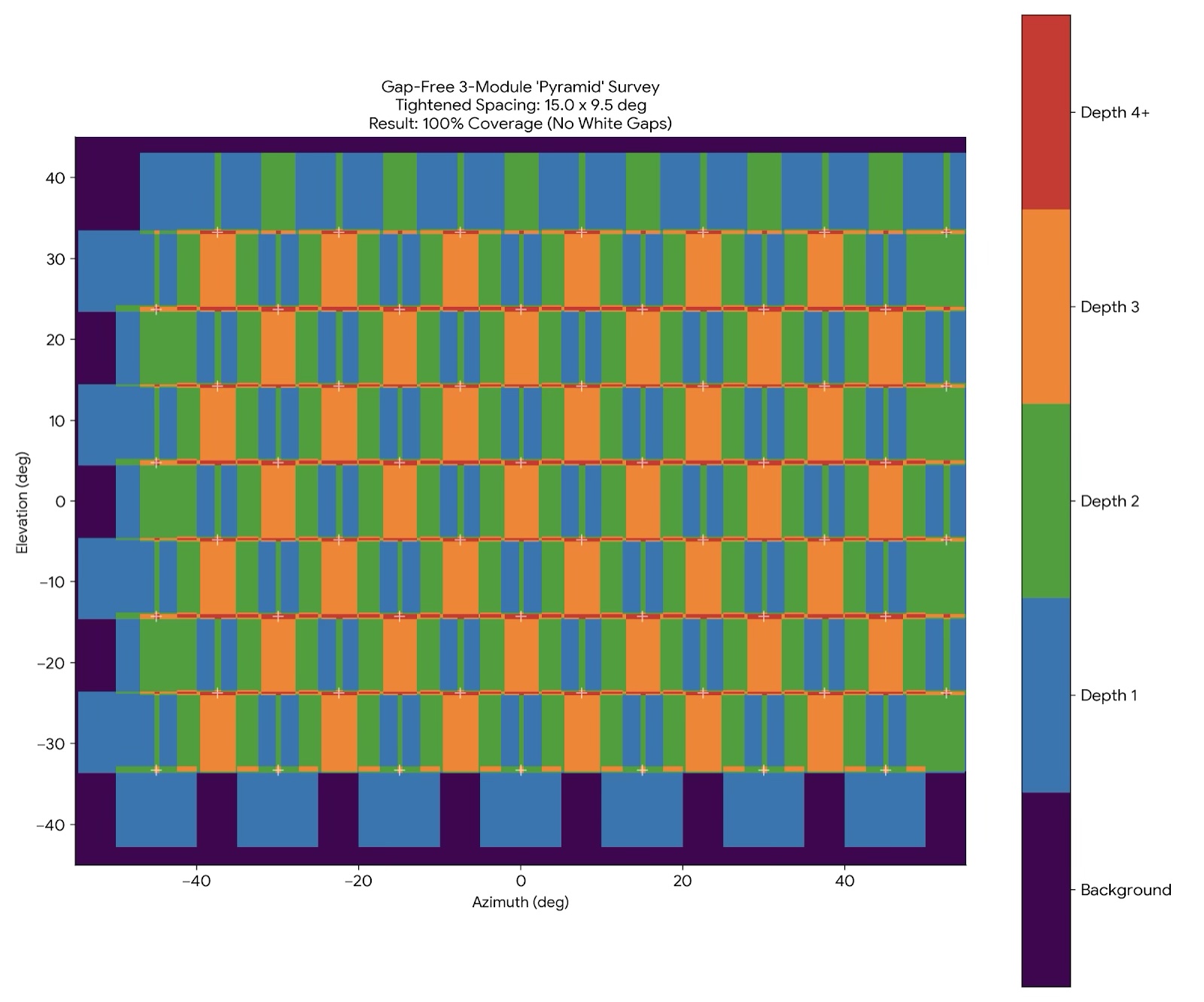}
\caption{A three-module T-shaped FoV formed with the same criteria used for the tri-petal (left) and an example tiling 6$\times$6 tiling (right). The tiling of a T-shaped FoV requires twice the survey times and produces a strongly structured non-uniform exposure time.}
\label{fig:tee}
\end{figure}

Other Cartesian square arrangements, such as a three-module T-shape, are appealing because they are simple. However, like the three-module L-shape, they drive the survey toward strongly varying exposure patterns requiring significantly longer survey times. They have tiling problems over the large areas required for LSST coverage and GW error boxes, which the tri-petal pattern effectively avoids (Figure~\ref{fig:tee}).  

Given the success of EP, a consideration of its observing strategy is in order. The FoV is similar to a Swiss cross, but with shorter arms, and all modules have similar rolls. The entire anti-Sun hemisphere is observed with 9 pointings over three 96.5 minute orbits. The pattern has a complex non-uniform response and the natural shift due to the Earth's motion smooths that response pattern over several days. The effective area is 2-3 cm$^2$ and typical exposures are $\sim1$ ks, so each complete hemisphere image is very shallow and less sensitive to fast transients.  The longer focal length allows XTRA/MOXI reach sources $\sim3$ times fainter per 1 ks exposure, allowing XTRA/MOXI to sample $\sim6$ times more Euclidean volume per unit sky area. XTRA/MOXI also extends the bandpass to 0.2 keV, providing critical sensitivity to the very soft emission from shock breakouts, TDEs, and soft GRB-like transients.

EP/WXT is optimized for rapid, shallow, broad soft X-ray sky coverage, with
EP/FXT providing deeper narrow-field follow-up of selected transients.  XTRA/MOXI
addresses a complementary trade: fewer wide-field modules with controlled
differential roll and overlap, so that the same focusing soft X-ray instrument
provides wide-area survey coverage, a high-depth survey floor, and a compact
pointed-observation core.

\section{Pointed observations and multimessenger follow-up}

For GRB and multimessenger follow-up, the operational implication is straightforward: the first pointing places the best available burst position, localization centroid, candidate host galaxy, or externally identified counterpart in the three-module core, while the surrounding bridge and petal regions retain coverage of nearby probability.  If the initial localization is larger than the compound field, subsequent offset pointings can tile the remaining probability according to the expected fading time and alert priority.

\section{Extension to persistent X-ray monitor arrays}
\label{sec:hydra}

Ground-based optical time-domain astronomy is moving toward persistent monitoring of very large fields, exemplified by Argus and its \(\sim8000\ \degsq\) instantaneous field at seconds-to-minutes cadence \cite{Law2021Argus,Law2022Pseudofocal,Corbett2022ArgusHDPS}.  The tri-petal geometry developed here offers a complementary route to a future persistent soft X-ray monitor: rather than simply increasing sky area, it uses overlapping fields of view to increase sensitivity, rotates the lobster-eye cross-arm PSF to reduce source confusion, and provides graceful degradation from module loss. A future Hydra-like mission---the name reflecting a many-headed architecture of coordinated wide-field soft X-ray telescope units rather than one monolithic telescope---would extend this idea to persistent, focusing coverage of a large selected anti-Sun field.

For XTRA, adjacent pointings build a multi-view exposure floor over time. In a persistent monitor array, the same hexagonal-knit lattice could instead be populated by physical tri-petal cells, making the XTRA survey pattern instantaneous. The design trade would then shift from producing a compact central overlap region for pointed observations to optimizing the offset and pitch for survey completeness, vignetting-weighted depth uniformity, localization, source-confusion mitigation, and robustness.  A representative Hydra-like implementation using EP/WXT-like \(f\simeq375\ {\rm mm}\) MPO units with \(\sim14.5^\circ\) fields would require of order 75 tri-petal cells, or \(\sim225\) telescope units, to cover 
$2\pi$ steradians at dense XTRA-like overlap depth; smaller cost-constrained implementations would trade sky area against overlap depth. These numbers are illustrative, since a real design must replace integer overlap with energy-dependent vignetting, PSF, detector background, detector gaps, source spectra, and onboard detection efficiency.

\section{Conclusions}

The design of wide-field micropore-optics soft X-ray instruments involves coupled trade-offs between single-module performance, system architecture, and survey geometry.  It was clear that modular designs provide scalability, redundancy, simpler repeated integration and calibration, and freedom to choose boresight offsets and roll angles. For reasons of schedule, mass, and budget we had chosen an architecture intermediate between a monolithic optic and the SIBEX solution of 24 modules of 2$\times$2 MPO facets; three or four modules with 5$\times$5 MPO facets resulting in a $\sim10^\circ\times\sim10^\circ$ FoV for each module. It was clear that one could overlap the modules to allow an instrument that was simultaneously capable of both deep pointed observations and shallow large-area survey observations. The question was then how to choose between three and four modules, and what observation strategy optimizes the rate at which transients are detected, and allows timely uniform coverage of large ($\sim10^3$ square degree) error boxes.

A four module architecture is locked into identical roll angles (Figure~\ref{fig:four}), which will align PSF arms in regions of overlap, while a three module architecture allows each module to have a different roll. Three module architectures with no differential roll tile very inefficiently and with strong variation in survey depth. In this work we have shown how these considerations lead naturally to a tri-petal field geometry. The differential roll in the tri-petal configuration is not an embellishment; it is the enabling step that turns square-format MPO modules into an effective hexagonal survey cell, which then allows uniform sky coverage over very large areas. While the four module (square) architecture provides larger instantaneous area and thus a faster survey, the tri-petal architecture allows a survey twice as deep with greater uniformity.  Conversely, the loss of one module of four requires difficult, inefficient tiling while the loss of one module of the tri-petal has a graceful alternative tiling. Compared with no-rotation or Cartesian alternatives, the tri-petal arrangement provides a better compromise between survey area, overlap depth, pointed sensitivity, temporal sampling, and module count.

The relative spacings of the three petals can be ``tuned''; we have chosen to maximize the survey area while still providing an overlap a few times larger than the expected PSF size; the overlap is $0.4^\circ$. Assuming an efficient serpentine scan, we can tune pitch, the size of the survey steps, in two different dimensions, in order to  maximize the uniformity of the resulting mosaic. We found that a $10^\circ$ step perpendicular to a side of the natural hexagon ``formed'' by the tri-petal, and a $8.6^\circ$ step in the perpendicular direction optimized coverage. This choice automatically provides multiple visits over an interval that can also be tuned; a property that is important for temporal studies. The repeated pointings interlock in a hexagonal lattice, turning gaps into additional exposure and producing a high-depth survey floor, allowing more sensitive detection of transients over a larger area.

Einstein Probe provided the flight-proven modular-MPO reference point.  XTRA/MOXI explores the complementary compact, overlap-optimized regime in which controlled module overlap supports both wide-area survey discovery and pointed transient characterization. The same principle can be scaled to Hydra-like persistent monitors for Argus-scale or larger fields, where many tri-petal cells operate simultaneously.

\section*{Code and Data Availability}

The results presented in this article are based on analytic instrument-geometry calculations, micropore-optics effective-area estimates, and survey-tiling simulations developed for the XTRA/MOXI trade study. No observational data or proprietary third-party datasets were analyzed. The numerical values and derived quantities needed to interpret the results are provided within the article. The Python code used to generate the survey-geometry simulations, overlap maps, and figures will be made publicly available in a GitHub repository prior to publication, and the final repository citation and DOI will be added to the reference list when available.

\section*{Acknowledgments}

The authors thank the XTRA study team for discussions that shaped the tri-petal survey-geometry trade study. We thank the MOXI instrument team for input on the micropore-optics module concept, focal-plane geometry, and survey-configuration trades. We also acknowledge discussions with Brendan O'Connor and  colleagues working on wide-field optical time-domain surveys, including Argus-like persistent-monitoring concepts, which motivated the extension of the tri-petal geometry to larger instantaneous X-ray arrays. 

\section*{Disclosures}

The authors declare there are no financial interests, commercial affiliations, or other potential conflicts of interest that have influenced the objectivity of this research or the writing of this paper. The authors used OpenAI ChatGPT and Google Gemini to assist with code generation, code debugging, language editing, grammar cleanup, and preparation of the cover letter. Representative prompts included requests to generate or revise Python code for survey-geometry simulations and figure generation, debug and improve analysis scripts, improve manuscript wording, draft the journal submission cover letter, and add the required large-language-model disclosure. All AI-assisted text and code were reviewed, edited, and validated by the authors. The scientific analysis, figures, interpretation, and conclusions are the responsibility of the authors.


\bibliographystyle{spiejour}
\bibliography{report_spieman_fixed-3_journalized}


\vspace{2ex}\noindent\textbf{Nicholas E. White} is a Research Professor of
Physics at George Washington University and a Fellow of the American Physical
Society and the American Astronomical Society.  His career has focused on
high-energy astrophysics and the development of space observatories.  He has
published more than 200 refereed papers on X-ray binaries, gamma-ray bursts,
stellar coronae, and black-hole accretion, and has played key roles in the
development and scientific use of the High Energy Astrophysics Research Center (HEASARC) and multiple international X-ray observatories.

\vspace{2ex}\noindent\textbf{Massimiliano Galeazzi} has worked for the last twenty-five years in the field of X-ray Astrophysics both in the experimental and data analysis fields. He was the PI of the DXL mission that successfully launched four times and was the first mission to successfully deploy micropore optics in space, and he is now the PI of the LXT mission, consisting of 4 co-aligned micropore optics telescopes coupled to CCD detectors. He was part of the LEXI mission that landed an X-ray telescope on the surface of the Moon in 2025, part of the Hitomi Team, and a member of the XMM-Newton Science Working Group. He is involved in data analysis and simulations from XMM-Newton, Chandra, ROSAT, eROSITA, and Suzaku for the study of Solar Wind Charge eXchange, the Diffuse X-ray Background, and the Warm-Hot Intergalactic Medium. He is a Professor and Chair of the Department of Physics at the University of Miami. 

\vspace{2ex}\noindent\textbf{Russell Roberts} is a third year graduate student at the University of Miami. His research focuses on the simulation and manufacturing of Micropore Optics and the design and optimization of future MPO based missions. He is also a member of the LXT sounding rocket missions, in charge of instrument performance, electrical interfaces, and ground support equipment. 

\vspace{2ex}\noindent\textbf{K. D. Kuntz} is Principal Research Scientist at Johns Hopkins University and a member of the XMM-Newton Guest Observer Facility at GSFC. His special interest is the soft diffuse X-ray background and the backgrounds, instrumental and otherwise, that must be characterized to isolate the cosmic X-ray background. His work with the XMM-Newton background produced some of the first studies of solar wind charge exchange. He has had been calibrating and simulating MPO since the 2012 flight of DXL.

\end{spacing}
\end{document}